\documentclass[hidelinks,onefignum,onetabnum]{siamart220329}

\usepackage{nicefrac}
\usepackage{multirow}
\usepackage{multicol}
\usepackage{array}
\usepackage{acronym}
\usepackage{colortbl}
\usepackage{amsfonts}
\usepackage{lipsum}
\usepackage{tikz}
\usepackage[english]{babel}
\usepackage{hyperref}
\usepackage{graphbox}
\usepackage{listings}
\usepackage{mathtools}
\usepackage{pgfplots}
\usepackage{subfloat}
\usepackage{xurl}
\usepackage{algorithm}
\usepackage[noend]{algpseudocode}
\usepackage{hyperref}
\usepackage{bm}
\usepackage{xspace}
\usepackage{wrapfig}
\usepackage{subfig}
\usepackage{soul}
\usepackage{booktabs}
\usepackage[range-phrase = \,--\,, range-units = single]{siunitx}
\usepackage{crefthe}
\hypersetup{
	colorlinks=true,
	linkcolor=blue,
	citecolor=blue,
	urlcolor=blue
}

\newtheorem{remark}{Remark}

\crefname{section}{Section}{Sections}
\crefname{subsection}{Section}{Sections}
\crefname{subsubsection}{Section}{Sections}
\crefname{lstlisting}{Listing}{Listings}

\usepackage{pgfplotstable}

\pgfplotstableset{ every head row/.style = { before row = \toprule
                                           , after row  = \midrule
                                           }
                 , every last row/.style = { after row  = \bottomrule }
                 , column type = S   
                 , string type       
                 , multicolumn names 
                 }

\acrodef{DoF}{degree of freedom}
\acrodefplural{DoF}{degrees of freedom}
\acrodef{DSL}{domain specific language}
\acrodef{FMG}{full multigrid}
\acrodef{PDE}{partial differential equation}
\acrodef{AST}{abstract syntax tree}
\acrodef{ECM}{execution-cache-memory model}
\acrodef{FMA}{fused multiply add}
\acrodef{LD}{load}
\acrodef{ST}{store}
\acrodef{CSE}{common subexpression elimination}
\acrodef{FFC}{FEniCS form compiler}
\acrodef{FE}{finite element}
\acrodef{RHS}{right-hand side}
\acrodef{FLOP}{floating point operation}

\usepackage{todonotes}

\definecolor{darkteakgreen}{RGB}{0, 128, 128}

\definecolor{someorange}{RGB}{226,135,67}

\usetikzlibrary{patterns}
\usetikzlibrary{arrows}
\usetikzlibrary{shapes}
\usetikzlibrary{external}
\usetikzlibrary{arrows.meta}
\usetikzlibrary{decorations.pathreplacing}
\usetikzlibrary{positioning}

\usepgfplotslibrary{fillbetween}

	\pgfplotsset{every axis/.style =%
  { font=\footnotesize
  , grid style = {line width=.2pt, draw=gray!20}
  , major grid style = {line width=.2pt,draw=gray!50}
  },
  colormap name = viridis,
  cycle multiindex* list = {
    [samples of colormap = 9]\nextlist
    mark list*\nextlist
  },
  /pgfplots/bar cycle list/.style={
    cycle multiindex* list = {
      [samples of colormap = 9]\nextlist
      mark list*\nextlist
      {fill=.!30}\nextlist
    },
  }
}

\pgfplotsset{roofline/.style =%
  { xlabel = {arithmetic intensity [\unit{FLOP/byte}]}
  , ylabel = {performance [\unit{GFLOP/s}]}
  , ytick = {1346.8,697.3,390.9,195.5}
  , yticklabels = {$\num{1347}$,$\num{697}$,$\num{391}$,$\num{195}$}
  , ymax = 2500
  , width = 0.45\linewidth
  , height = 0.3\linewidth
  , grid = both
  , mark options = {scale = 1.5}
  }%
}

\newcommand{\drawroofline}[1][]{%
  \newif\ifltwo
  \newif\ifavxfma
  \newif\ifavx
  \newif\iffma
  \newif\ifpeak
  \pgfkeys{/roofline/.cd
          , xmax/.initial=40
          , xmax/.get=\xmax
          , xmax/.store in=\xmax
          , xmin/.initial=0.55
          , xmin/.get=\xmin
          , xmin/.store in=\xmin
          , angle/.initial=36
          , angle/.get=\angle
          , angle/.store in=\angle
          , ltwo/.is if=ltwo
          , ltwo=false
          , avxfma/.is if=avxfma
          , avxfma=true
          , avx/.is if=avx
          , avx=true
          , fma/.is if=fma
          , fma=true
          , peak/.is if=peak
          , peak=true
          , #1
          }

  \ifltwo
    \addplot[domain=0.560:\xmax] {1346.83896}; 
    \addplot[domain=0.290:\xmax] { 697.28972}; 
    \addplot[domain=0.162:\xmax] { 390.91116}; 
    \addplot[domain=.0812:\xmax] { 195.45558}; 
    \addplot[domain=.0406:0.560] {2406.08908*x};
  \else
    \addplot[domain=7.590:\xmax] {1346.83896}; 
    \addplot[domain=3.929:\xmax] { 697.28972}; 
    \addplot[domain=2.203:\xmax] { 390.91116}; 
    \addplot[domain=1.101:\xmax] { 195.45558}; 
    \addplot[domain=\xmin:7.590] { 177.46063*x};
  \fi
  \pgfplotsset{cycle list shift = -5}

  \begin{scope}[yshift=-2]
    \ifavxfma
      \node[anchor=south east] at (axis cs:\xmax,1347) {avx fma};
    \fi
    \ifavx
      \node[anchor=south east] at (axis cs:\xmax, 697) {avx};
    \fi
    \iffma
      \node[anchor=south east] at (axis cs:\xmax, 391) {fma};
    \fi
    \ifpeak
      \node[anchor=south east] at (axis cs:\xmax, 195) {scalar};
    \fi
    \ifltwo
      \node[anchor=south,rotate=\angle] at (axis cs:.18,391) {L2 bw \qty{2406}{GB/s}};
    \else
      \node[anchor=south,rotate=\angle] at (axis cs:2.0,340) {mem bw \qty{177}{GB/s}};
    \fi
  \end{scope}
}

\pgfplotsset{performance/.style =%
  { xbar
  , bar shift = 0pt
  , bar width = 1.5ex
  , xlabel = {performance [MDoF/s]}
  , minor x tick num = 4
  , xmin = 0
  , enlarge y limits = 0.1
  , width = 0.38\linewidth
  , height = 0.3\linewidth
  , xmajorgrids = true
  , xminorgrids = true
  , mark options = {scale = 1.5, fill = .}
  }%
}

\tikzset{
  n/.style =%
    { font = \tiny
    , fill = white
    , inner sep = 0.5pt
    },
  a/.style =%
    { -{Latex[length=4pt]}
    , shorten <=3pt
    , shorten >=3pt
    },
 mark options = {scale = 1.5}
}

\lstdefinestyle{commandline}{
    basicstyle=\small\ttfamily,
    backgroundcolor=\color{gray!10},
    frame=tb,
    numbers=none,
    showstringspaces=false,
}

\lstdefinestyle{pythonstyle}{
    language=Python,
    basicstyle=\small\ttfamily,
    keywordstyle=\color{blue}\bfseries,
    stringstyle=\color{green!50!black},
    commentstyle=\color{gray}\itshape,
    numbers=left,
    numberstyle=\tiny\color{gray},
    stepnumber=1,
    numbersep=5pt,
    backgroundcolor=\color{gray!5},
    frame=tb,
    showspaces=false,
    showstringspaces=false,
}

\counterwithout{figure}{section}
\counterwithout{table}{section}
\counterwithout{algorithm}{section}
\numberwithin{equation}{section}

\newcommand{\ourtitle}[0]{Code Generation and Performance Engineering for \\ Matrix-Free Finite Element Methods on \\ Hybrid Tetrahedral Grids}
\newcommand{\ourshorttitle}[0]{Matrix-Free Finite Element Methods on Hybrid Tetrahedral Grids}

\newcommand{\hyteg}{\textsc{HyTeG}\xspace}

\newcommand{\fogshort}{\textsc{HOG}\xspace}
\newcommand{\sympy}{\textsc{Sympy}\xspace}

\newcommand{\pystencils}{\textsc{Pystencils}\xspace}
\newcommand{\cpp}{C++\xspace}
\newcommand{\likwid}{LIKWID\xspace}
\newcommand{\quadpy}{\textsc{quadpy}\xspace}
\newcommand{\basix}{\textsc{basix}\xspace}
\newcommand{\fenics}{FEniCS\xspace}

\makeatletter
\newcommand{\replunderscores}[1]{\expandafter\@repl@underscores#1_\relax}

\def\@repl@underscores#1_#2\relax{%
    \ifx \relax #2\relax
        #1%
    \else
        #1%
        \textunderscore
        \@repl@underscores#2\relax
    \fi
}
\makeatother
\newcommand{\inst}[1]{\replunderscores{#1}\,\ref{pgf:#1}}

\newcommand{\genOp}[0]{A}
\newcommand{\genLocOp}[0]{A_{T}}
\newcommand{\genLocMicroOp}[0]{A_{T_m}}
\newcommand{\genVec}[1]{\mathbf{#1}}

\newcommand{\grid}[1]{\mathcal{T}(#1)}

\newcommand{\projOp}[2]{R_{#1}^{#2}}

\newcommand{\intMicroElem}{\int_{T_m}}

\newcommand{\basis}{\phi}
\newcommand{\trialbasis}{\phi}
\newcommand{\testbasis}{\psi}

\newcommand{\locMat}[3]{\left[ #1 \right]_{j \in #2, i \in #3}}

\newcommand{\macro}{T_M}
\newcommand{\macroGrid}[1]{\mathcal{T}_M(#1)}
\newcommand{\microGrid}[1]{\mathcal{T}_m(#1)}
\newcommand{\micro}{T_m}
\newcommand{\microsInMacro}[2]{\macro(#1, #2)}

\newcommand{\trialspacena}{\mathcal{V}}

\newcommand{\testspacena}{\mathcal{W}}
\newcommand{\coeffspacena}{\mathcal{K}}
\newcommand{\quadPoly}{\mathcal{P}_2}
\newcommand{\linPoly}{\mathcal{P}_1}
\newcommand{\nedelec}{\mathcal{ND}_1}
\newcommand{\curl}{\mathbf{curl} \ }

\newcommand{\iset}[1]{\mathcal{I}^{#1}}

\newcommand{\orientation}{
\begin{tikzpicture}[scale=0.22]
\fill[gray] (0,0) -- (1,0) -- (0.5,1) -- cycle;
\draw[black] (0,0) -- (1,0) -- (0.5,1) -- cycle; 
\end{tikzpicture}
}

\newcommand{\RNum}[1]{\uppercase\expandafter{\romannumeral #1\relax}}
\newcommand{\whiteup}{
\begin{tikzpicture}[scale=0.22]
	\clip (0,0) rectangle (1,1);
\fill[white] (0,0) -- (1,0) -- (0.5,1) -- cycle;
\draw[black] (0,0) -- (1,0) -- (0.5,1) -- cycle;
\draw[black] (0.5,0) -- (0.5,1) -- cycle; 
\end{tikzpicture}
}
\newcommand{\whitedown}{
	\begin{tikzpicture}[scale=0.22]
		\clip (0,0) rectangle (1,1);
	\fill[white] (0,1) -- (1,1) -- (0.5,0) -- cycle;
	\draw[black] (0,1) -- (1,1) -- (0.5,0) -- cycle; 
	\draw[black] (0.5,0) -- (0.5,1) -- cycle; 
	\end{tikzpicture}
}

\newcommand{\blueup}{
\begin{tikzpicture}[scale=0.22]
\fill[white] (0,0) -- (1,0) -- (0.5,1) -- cycle;
\draw[black] (0,0) -- (1,0) -- (0.5,1) -- cycle; 
\draw[black] (0.35,0) -- (0.35,0.65); 
\draw[black] (0.65,0) -- (0.65,0.65); 
\end{tikzpicture}
}
\newcommand{\bluedown}{
\begin{tikzpicture}[scale=0.22]
\fill[white] (0,1) -- (1,1) -- (0.5,0) -- cycle;
\draw[black] (0,1) -- (1,1) -- (0.5,0) -- cycle; 
\draw[black] (0.35,0.35) -- (0.35,1); 
\draw[black] (0.65,0.35) -- (0.65,1); 
\end{tikzpicture}
}
\newcommand{\greenup}{
	\begin{tikzpicture}[scale=0.22]
	\fill[white] (0,0) -- (1,0) -- (0.5,1) -- cycle;
	\draw[black] (0,0) -- (1,0) -- (0.5,1) -- cycle; 
	\draw[black] (0.5,0) -- (0.5,1) ; 
	\draw[black] (0.3,0) -- (0.3,0.6); 
	\draw[black] (0.7,0) -- (0.7,0.6); 
	\end{tikzpicture}
}
\newcommand{\greendown}{
	\begin{tikzpicture}[scale=0.22]
	\fill[white] (0,1) -- (1,1) -- (0.5,0) -- cycle;
	\draw[black] (0,1) -- (1,1) -- (0.5,0) -- cycle;
	\draw[black] (0.5,0) -- (0.5,1) ; 
	\draw[black] (0.3,0.35) -- (0.3,1); 
	\draw[black] (0.7,0.35) -- (0.7,1);  
	\end{tikzpicture}
}
\newcommand{\allOrientations}{\{ \whiteup, \whitedown, \blueup, \bluedown, \greenup, \greendown \}}
\newcommand{\position}{x,y,z}

\headers{\ourshorttitle}{Böhm, Bauer, Kohl, Alappat, Thönnes, Mohr, Köstler, Rüde}

\title{\ourtitle}
\author{
	Fabian Böhm\footnotemark[2]
	\thanks{Erlangen National High Performance Computing Center (NHR@FAU), Erlangen, Germany}
	\and
	Daniel Bauer\footnotemark[2] \footnotemark[1]
	\and
	Nils Kohl\footnotemark[3]
	\and
	Christie Alappat\footnotemark[1]
	\and\newline
	Dominik Thönnes\thanks{Friedrich-Alexander-Universität Erlangen-Nürnberg (FAU), Erlangen, Germany\newline        
		(\{fabian.boehm, daniel.j.bauer, christie.alappat, dominik.thoennes, harald.koestler, \newline ulrich.ruede\}@fau.de).} 
	\and
	Marcus Mohr\thanks{Dept.~of Earth and Environmental Sciences, Ludwig-Maximilians-Universität München (LMU), Munich, Germany 
		(\{nils.kohl, \mbox{marcus.mohr}\}@lmu.de).}
	\and
	Harald Köstler\footnotemark[2] \footnotemark[1]
	\and
	Ulrich Rüde\footnotemark[2] \thanks{Centre Européen de Recherche et de Formation Avancée en Calcul Scientifique (CERFACS), Toulouse, France. }
}

\begin{document}
\maketitle
\begin{abstract}
	This paper introduces a code generator designed for node-level optimized, extreme-scalable,
	matrix-free finite element operators on hybrid tetrahedral grids. It optimizes
	the local evaluation of bilinear forms through various techniques including
	tabulation, relocation    of loop invariants, and inter-element vectorization -
	implemented as transformations of an abstract syntax tree. A key contribution
	is the development, analysis, and generation of efficient loop patterns
	that leverage the local structure of the underlying tetrahedral grid. These
	significantly enhance cache locality and arithmetic intensity, mitigating
	bandwidth-pressure associated with compute-sparse, low-order operators.
	The paper demonstrates the generator's capabilities through a comprehensive
	educational cycle of performance analysis, bottleneck identification,
	and emission of dedicated optimizations. For three differential operators
	($-\Delta$, $-\nabla \cdot (k(\mathbf{x})\, \nabla\,)$, $\alpha(\mathbf{x})\,
	\mathbf{curl}\ \mathbf{curl} + \beta(\mathbf{x}) $), we determine the set
	of most effective optimizations. Applied by the generator, they result in
	speed-ups of up to 58$\times$ compared to reference implementations. Detailed
	node-level performance analysis yields matrix-free operators with a throughput
	of 1.3 to 2.1 GDoF/s, achieving up to \qty{62}{\%} peak performance on a 36-core
	Intel Ice Lake socket. Finally, the solution of the curl-curl problem with more
	than a trillion ($ 10^{12}$) degrees of freedom on \num{21504} processes in
	less than \num{50} seconds demonstrates the generated operators' performance and
	extreme-scalability as part of a full multigrid solver.
\end{abstract}



\overfullrule=0pt

\section{Introduction}
\label{sec:Introduction}

Matrix-free finite element methods~\cite{Brown:2010:JSC,
Kohl:2022:TextbookEfficiencyMassively,
Kronbichler:2012:CellBasedOA,May:2015:ScalableMatrixfreeMultigrid,
Rudi:2015:ExtremeScaleImplicit}
address two main limitations
faced by conventional approaches that follow the assemble-solve cycle.
Matrix-vector operations using standard sparse storage formats are typically
bandwidth-limited on state-of-the-art
architectures~\cite{Kronbichler:2018:PerformanceComparisonContinuous}.
The characteristic machine balance, i.e., the ratio of memory bandwidth (B/s)
to performance (FLOP/s) of current hardware favors on-the-fly evaluation
that reduces bandwidth-pressure at the cost of additional arithmetic
operations~\cite{Kirby:2018:SolverComposition,
Kronbichler:2019:FastMatrixFreeEvaluation,
Kronbichler:2018:PerformanceComparisonContinuous,
May:2014:HPCLithospheric}.
Secondly, the available memory typically limits the spatial resolution if the
entire matrix has to be assembled and stored.
An example from the geosciences illustrates the latter issue.
The simulation of convection in the Earth's
mantle with a global resolution of \qty{1}{km} requires about a trillion ($10^{12}$)
elements~\cite{Bauer:2020:TerraNeoMantleConvection}.
Under the simplifying assumption that the underlying discretization yields one
degree of freedom per element
and an operator with a 7-point stencil,
we end up with a memory requirement of
\begin{align}\label{eq:mem-requirement-estimate}
	\underbrace{7}_{\text{non-zeros per row}} \cdot \underbrace{10^{12}}_{\text{matrix rows}} \cdot \underbrace{\qty{8}{B}}_{\text{double precision}} = \qty{56}{TB}
\end{align}
for the system matrix.
This estimate is extremely optimistic.
The number of degrees of freedom and the stencil size
are typically much larger
and the overhead required to store the indexing data structure of the sparse
matrix format is omitted here.
Expecting a memory requirement one order of magnitude higher than \cref{eq:mem-requirement-estimate},
storing the system matrix becomes infeasible, even on the majority of the largest available supercomputers.
Since most iterative linear solvers only require the results of
matrix-vector operations but no explicit access to the matrix entries,
\emph{matrix-free} methods enable the solution of linear systems with trillions
($10^{12}$) of
unknowns~\cite{Gmeiner:2016:QuantitativePerformanceStudy,
Kohl:2022:TextbookEfficiencyMassively}
that could not be realized with standard sparse assembly.

However, the implementation of efficient matrix-free methods is challenging
since the matrix-free execution requires that the discretization and the linear
solvers are coupled more tightly.
Performing a matrix-free matrix-vector multiplication requires
information about the underlying mesh, finite element spaces
and the differential operators.
At the same time, having this knowledge enables further domain-specific optimizations that cannot be exploited
using standard sparse linear algebra.
The extensive range of combinations of differential operators, finite element spaces, application-dependent
optimizations and target platforms renders the manual implementation and optimization of compute kernels a daunting
task.
Not only because of the amount of code that needs to be developed, maintained and tested, but also because expertise
from both numerical mathematics and performance engineering is crucial to leverage the full potential of the underlying
hardware. 

\subsection{Contribution}

This paper presents the \hyteg Operator Generator\footnote{\url{https://i10git.cs.fau.de/hyteg/hog}} (\fogshort) -- a unified
pipeline that realizes the automated generation of
matrix-free compute kernels from a symbolic description of a differential
operator and respective finite element spaces.
The compute kernels are tailored to block-structured, hybrid tetrahedral grids
that enable direct addressing of unknowns via implicit, analytical index
mappings for fast, contiguous and predictable memory access.
Specifically, they are integrated into the \hyteg finite element framework\footnote{\url{https://i10git.cs.fau.de/hyteg/hyteg}}
\cite{Kohl:2023:FundamentalDataStructures,Kohl:2019:HyTeGFiniteelementSoftware}.

The main focus is put on the optimization of the matrix-free matrix-vector
product $y \gets Ax$, where $A$ is the system matrix stemming from a finite
element discretization of a \ac{PDE}. 
To demonstrate the flexibility of the code generator, we analyze bilinear forms
that arise from the weak formulation of three different differential
operators ($-\Delta$, $-\nabla \cdot (k(\mathbf{x})\, \nabla\,)$, $\alpha(\mathbf{x})\, \curl \curl\! + \beta(\mathbf{x}) $)
for different types of finite element 
spaces (linear and quadratic Lagrange, first order Nédélec).

A central
contribution of this paper is the in-depth, step-by-step performance
analysis guiding the generation of optimized compute kernels through
resource-based performance models~\cite{Hager:2010:IntroductionHighPerformance}.
Optimizations include identification of loop invariants in the \ac{AST}, 
inter-element vectorization, tabulation of factors of the weak form and several others. Tailored loop patterns that exploit the underlying structure of the grid to enhance cache locality are presented and analyzed.
We evaluate the efficiency of the applied optimizations via
the roofline performance model~\cite{Williams:2009:Roofline}, layer condition
analysis~\cite{Hammer:2017:kernkraft,Stengel:2014:StencilPerfBottlenecks},
and analytical bounds for the memory traffic. For a range of weak forms, the set of most effective optimizations is identified from a larger pool of optimizations.
Our analysis qualitatively and quantitatively demonstrates what limitations
have to be overcome to achieve high node-level performance.

\begin{figure}
    \begin{tikzpicture}
      \begin{axis}
        [ performance
        , symbolic y coords = {T,C,I,U,V, S, R}
        , ytick = { R, S, V, U,I,C,T}
        , yticklabels = {reference, symmetry, {inter-element  vectorization}, under-integration, move loop invariants, cubes loop strategy, tabulation } 
        , xlabel = {single socket performance [MDoF/s]}
       , width = 0.6\linewidth
        ]
        \addplot[color of colormap={0},fill=.!30,mark=*] coordinates {(22.3,R)}; 
         \draw[a, out=0, in=90] (axis cs:22.310554513795875,R) to node[n, anchor=south west, xshift=0.8cm, yshift=-0.1cm]{\normalsize total: $\num{58}\times$} (axis cs:1300,T);
        \addplot[color of colormap={100},fill=.!30,mark=*] coordinates {(37.3,S)};  
         \addplot[color of colormap={200},fill=.!30,mark=*] coordinates {(120,V)};
         \addplot[color of colormap={300},fill=.!30,mark=*] coordinates {(463,U)};
         \addplot[color of colormap={400},fill=.!30,mark=*] coordinates {(866,I)};
         \addplot[color of colormap={500},fill=.!30,mark=*] coordinates {(1212,C)};
          \addplot[color of colormap={600},fill=.!30,mark=*] coordinates {(1300,T)};
           \draw[a] (axis cs:22.310554513795875,R) to node[n, anchor=south west, xshift=0.2cm, yshift=-0.1cm]{$\num{1.7}\times$} (axis cs:37.278759510666866,S);
             \draw[a, bend left] (axis cs:37.278759510666866,S) to node[n, anchor=south west, xshift=0.2cm, yshift=-0.1cm]{$\num{3.2}\times$} (axis cs:120.42962516572254,V);
          \draw[a, bend left] (axis cs:120.42962516572254,V) to node[n, anchor=south west, xshift=0.1cm]{$\num{3.8}\times$} (axis cs:463.6341372912801,U);
          \draw[a, bend left] (axis cs:463.6341372912801,U) to node[n, anchor=south west, xshift=0.1cm]{$\num{1.9}\times$} (axis cs:866.7648742411102,I);
          \draw[a, bend left] (axis cs:866.7648742411102,I) to node[n, anchor=south west, xshift=0.1cm]{$\num{1.4}\times$} (axis cs:1212.4271844660193,C);
          \draw[a, bend left] (axis cs:1212.4271844660193,C) to node[n, anchor=south west, xshift=0.16cm]{$\num{1.1}\times$} (axis cs:1294.7398963730568,T);
      \end{axis}
    \end{tikzpicture}
    \vspace{-0.3cm} 
  \caption{
  	Performance of generated matrix-free matrix-vector multiplication
  	kernels for a variable-coefficient diffusion operator
  	$-\nabla \cdot (k(\mathbf{x})\, \nabla\,)$ discretized with
  	quadratic conforming elements after application of individual performance
  	optimizations.
  	The final operator exhibits an accumulated speed-up of 58$\times$.
  	Detailed discussion in \cref{sec:PerfEng}.}
  \label{fig:speedup_ad}
    \vspace{-0.75cm}
\end{figure}
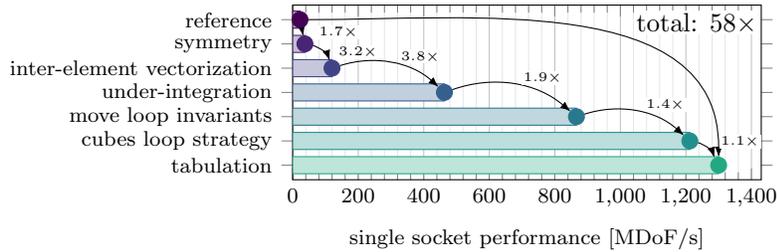

Concretely, the generated operators achieve a single-socket (36 cores)
throughput of \qtyrange{1.3}{2.1}{GDoF/s} on an Intel Xeon
IceLake architecture.
Thoroughly exploiting the machine, the operators reach up to \qty{62}{\%} of the machine's
peak performance and lie close to the machine balance due to optimizations
targeting the arithmetic intensity.

\cref{fig:speedup_ad} summarizes the effect of the individual optimizations
available in the code generator for a variable-coefficient diffusion operator
discretized by quadratic, conforming finite elements.
Compared to a reference implementation, we achieve an accumulated speed-up of
$58\times$ in this example.
\cref{sec:PerfEng} discusses more results and details of the analysis.

Finally, \hyteg's full multigrid solver equipped with the generated matrix-free
kernels demonstrates extreme scalability by solving a curl-curl 
problem
with
more than a trillion ($>10^{12}$) \acp{DoF} on \num{21504} processes in less
than $50$ seconds.

\subsection{Related Work}

Automated code generation has been successfully applied to accelerate the development
of \ac{FE} based applications for several years and in many projects, most prominently via the FEniCS
project~\cite{Alnaes:2015:FEniCSProjectVersion}.
FEniCS generates code for fast global matrix assembly from a weak \ac{PDE}
definition in a dedicated, \ac{DSL} and using the
\ac{FFC}~\cite{Kirby:2006:FFC}.
The resulting assembled linear system is solved with black-box solvers from
the PETSc package~\cite{PETSc:1997:Efficient}.
The present approach is similar but focuses on the generation of
\emph{matrix-free} \ac{FE} methods, with a co-design of discretization,
matrix-free operator and geometric multigrid solver.
This is in contrast to the discretization-solver split implemented in
FEniCS and offers the option to exploit the type of discretization in the solver.
For this work, extreme scalability and high node-level performance are a first
order design goal, which goes to the cost of having a narrower range of
discretizations and \ac{PDE}s than FEniCS.

A similar approach is realized through the Firedrake
project~\cite{Ham:2023:Firedrake}.
Like FEniCS, it uses a \ac{DSL} to define the weak form
and PETSc as a solver backend.
Firedrake also supports matrix-free evaluation of matrix-vector products within
Krylov solvers~\cite{Kirby:2018:SolverComposition}, with assembled matrices
in the preconditioner.

The ExaStencils~\cite{RuedeKoestler:2020:ExaStencils} framework offers code
generation of whole programs. It implements a multi-layered language approach
where each of the four layers provides
a separate \ac{DSL} that is tailored for domain experts from different communities.
Problems can be generated from a textbook-like definition, where the discretization,
solver and parallelization are generated by the framework.

\subsection{Structure}

The remaining article is structured as follows.
The domain partitioning follows the concept of hybrid
tetrahedral grids and is introduced in \cref{sec:HyTeG}.
\Cref{sec:FullOperatorGeneration} describes the architecture of the code
generator.
\Cref{sec:MatrixFree} defines the finite element setting and the
matrix-free application of an \ac{FE} operator.
\Cref{sec:LoopStrategies} presents loop strategies and discusses cache locality.
\Cref{sec:genOpts} summarizes the implemented performance optimizations.
\Cref{sec:PerfEng} proceeds with the performance analysis and optimization of
the generated kernels for different differential operators.
Finally, \cref{sec:ExtremeScale} demonstrates the scalability of the generated operators in a matrix-free multigrid solver applied to the 
curl-curl problem.

\section{Hybrid Tetrahedral Grids}

\label{sec:HyTeG}

The optimizations presented in this paper are tailored towards a
block-structured domain partitioning.
Specifically, we embed the generated compute kernels into the \hyteg framework
for matrix-free, large-scale \ac{FE}
simulations~\cite{Kohl:2019:HyTeGFiniteelementSoftware}.
\hyteg is based on the concept of hierarchical hybrid
grids (HHG)~\cite{Bergen:2004:HierarchicalHybridGrids}.
The domain is approximated by an unstructured tetrahdral coarse grid that is
uniformly refined according to~\cite{Bey:1995:TetrahedralGridRefinement,Kohl:2023:FundamentalDataStructures}.
We refer to the elements of the unstructured coarse grid as macro-tetrahedra
and to the tetrahedra that emerge from the refinement as micro-tetrahedra.

\begin{figure}[t]
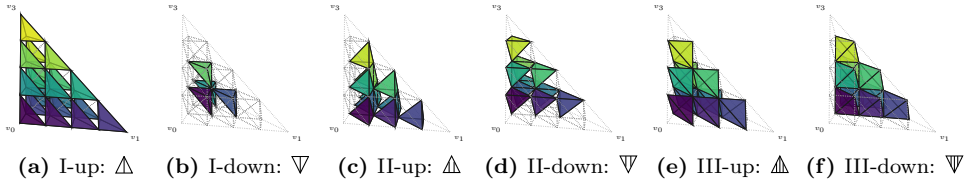

	\footnotesize
	\captionsetup[subfigure]{margin=0pt}
	\centering
	\vspace{-10pt}
	\subfloat[][\RNum{1}-up:\whiteup]{
		\resizebox{0.15\linewidth}{!}{\input{figures/micro_cell_types_width_5_cell_type_WHITE_UP}}
	}
	\subfloat[][\RNum{1}-down:\whitedown]{
		\resizebox{0.15\linewidth}{!}{\input{figures/micro_cell_types_width_5_cell_type_WHITE_DOWN}}
	}
	\subfloat[][\RNum{2}-up:\blueup]{
		\resizebox{0.15\linewidth}{!}{\input{figures/micro_cell_types_width_5_cell_type_BLUE_UP}}
	}
	\subfloat[][\RNum{2}-down:\bluedown]{
		\resizebox{0.15\linewidth}{!}{\input{figures/micro_cell_types_width_5_cell_type_BLUE_DOWN}}
	}
	\subfloat[][\RNum{3}-up:\greenup]{
		\resizebox{0.15\linewidth}{!}{\input{figures/micro_cell_types_width_5_cell_type_GREEN_UP}}
	}
	\subfloat[][\RNum{3}-down:\greendown]{
		\resizebox{0.15\linewidth}{!}{\input{figures/micro_cell_types_width_5_cell_type_GREEN_DOWN}}
	}
\vspace{-0.2cm}
	\caption{Six types of micro-elements
	with different orientations in space, illustrated on refinement level
		$\ell = 2$. Each group is denoted by a unique symbol to streamline the presentation of algorithms. Combined,
		they constitute the complete macro tetrahedron. Micros of a certain orientation are, as visible, translation invariant. Depending on the orientation, the macro-tetrahedron fits a differing number of micro-elements, e.g., four $\protect \whiteup$-elements against only two $\protect \whitedown$-elements in the longest row.
		See \cite{Kohl:2023:FundamentalDataStructures} for details.}
	\label{fig:hyteg-micro-cells}
	\vspace{-20pt}
\end{figure}

The resulting grid is fully structured within each macro-element, and each
new tetrahedron is identical up to translation to one of six reference
tetrahedra.
See~\cref{fig:hyteg-micro-cells} for an illustration of two times refined
macro-tetrahedra and the categorization of the arising micro-tetrahedra.
This local structure is heavily exploited by the compute kernels through
implicit indexing of the \ac{FE} data structures without indirections
or additional bookkeeping of connectivities.
Consecutive, direct memory access is crucial for the
high performance compute kernels presented in this paper.
The article~\cite{Kohl:2023:FundamentalDataStructures} provides a
detailed description of the refinement procedure and the resulting indexing
schemes.

An extreme-scalable data structure is constructed through the distribution of
the macro-primitives among parallel processes.
The uniform refinement enables the construction of matrix-free geometric
multigrid solvers by design, and thus the construction of asymptotically optimal
matrix-free full multigrid solvers that are essential to solve PDEs at the
extreme scale~\cite{Kohl:2022:TextbookEfficiencyMassively}.
The performance and scalability of this approach was
demonstrated in a series of articles, see
also~\cite{Bauer:2020:TerraNeoMantleConvection,
	Gmeiner:2015:PerformanceScalabilityHierarchical,
	Kohl:2023:FundamentalDataStructures,
	Kohl:2022:SISC,
	Thoennes:2023:ModelBasePerfAnalysisHyteg}.
Specifically, the solution of saddle point systems with more than ten
trillion ($10^{13}$) unknowns on hundreds of thousands of processes could be
realized due to the matrix-free
implementation~\cite{Gmeiner:2016:QuantitativePerformanceStudy}.

\section{Code Generation for FEM on Hybrid Tetrahedral Grids}
\label{sec:FullOperatorGeneration}

\begin{figure}[t]
	\footnotesize
	\centering
	\includegraphics[width=\textwidth, trim=0 10cm 10cm 2cm, clip]{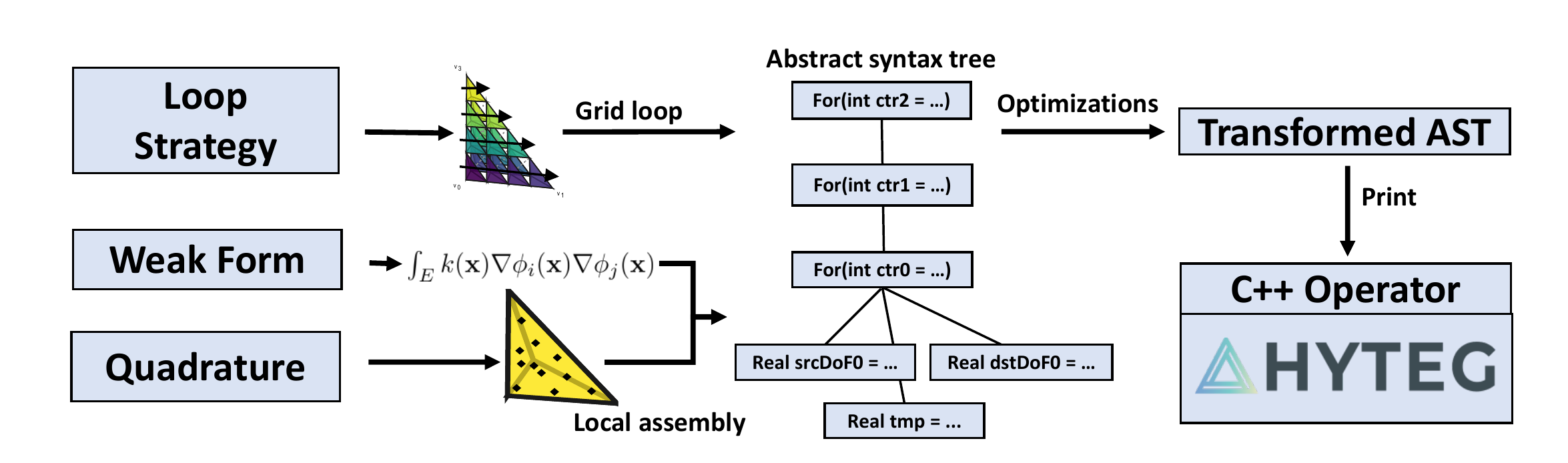}
	\vspace{-15pt}
	\caption{Code generation pipeline of the \fogshort. A scalable, matrix-free
		operator for hybrid tetrahedral grids is generated and optimized from
		simple input parameters. The resulting kernel is embedded into the
		\hyteg \ac{FE} framework \cite{Kohl:2019:HyTeGFiniteelementSoftware,Kohl:2023:FundamentalDataStructures}.}
	\label{fig:fogpipeline}
\vspace{-1cm}
\end{figure}

\fogshort implements a unified
pipeline to generate efficient \emph{matrix-free} \ac{FE} compute
kernels on block-structured tetrahedral grids.
Similar to the FEniCS-approach, it automatically generates kernels from the
symbolic description of a weak form and several other parameters, such as the
quadrature rule, \ac{FE} spaces and optimizations selected.
Specifically, the \fogshort takes a weak form as input in the shape of a
\sympy~\cite{Meurer:2017:Sympy} symbolic expression.
The quadrature points and weights are supplied by the \quadpy library
\cite{Schloemer:2021:Quadpy}.
An abstract syntax tree (AST) is constructed from the inputs using AST node
classes from \pystencils~\cite{Bauer:2019:Pystencils}, a library for the
generation of stencil codes.
Optimizations are applied to the \ac{AST}, before \cpp code is printed that will be called by the
\hyteg backend.
\Cref{fig:fogpipeline} visualizes the code generation pipeline.

The \fogshort generates kernels that execute an operation
on a single refined macro-primitive.
Different operations are supported, including matrix-free matrix-vector
multiplication and assembly of matrix diagonals, and new operations can
easily be added.
The generated code includes not only assembly of the element matrices on a
single element, but also encompasses the loop over the local subdomain.
The regular structure within a macro tetrahedron enables generation of
specially adapted loop strategies, which significantly improve memory access
patterns and, thus, the operator's performance.
Inter-element vectorization~\cite{Sun:2019:Vect}, i.e., computing multiple
elements in parallel using vectorized instructions, is a second central
optimization.

The \fogshort applies a wide range of efficient, state-of-the-art
quadrature rules for simplices, like the Xiao-Gimbutas
rules~\cite{Xiao:2010:Quad} and can be configured to under-integrate, when
applicable.
If reasonable, quadrature-free kernels can also be generated.
Additionally, the \fogshort applies established
optimization techniques such
as loop fusion, tree- or polynomial-based
\ac{CSE}~\cite{Hosangadi:2006:polyCSE}, exploitation of
symmetry and tabulation~\cite{Kirby:2006:FFC}, all adapted to hybrid
tetrahedral grids.

The generated operator uses \hyteg's MPI communication routines, providing the
desired scalability.
\hyteg's geometric multigrid and Krylov solvers then employ the generated
operators in smoothers, residual computations and coarse grid solves.

\section{Matrix-Free Finite Elements}

\label{sec:MatrixFree}

We consider the solution of linear systems of the form
$\genOp \genVec{v} = \genVec{f}$ arising from the discretization of linear
elliptic PDEs with the finite element method~\cite{Brenner:2008:FE,
Ciarlet:2002:FE} subject to
a triangulation $\grid{\Omega}$ of the domain $\Omega$ and finite dimensional
trial and test spaces $\trialspacena = \langle \trialbasis^{i} \rangle_{i\in \iset{\trialspacena}}$
and $\testspacena = \langle \testbasis^{j} \rangle_{i\in \iset{\testspacena}}$ with associated
basis functions $\trialbasis^i$ and $\testbasis^j$.

Most iterative linear solvers such as multigrid and Krylov methods do not
require access to the entries of $A$.
It is sufficient to provide a method to compute the \emph{result} of the
application of $A$ to a vector, which can be implemented via \emph{matrix-free}
kernels, enabling a solution of the linear system without the need to explicitly
form $A$.

With an integrand $G$ that originates from the weak formulation of the PDE,
we have that
\begin{align}
	\label{eq:matrixfreeop}
	\genOp =  \left[ \int_{\grid{\Omega}} G(\genVec{x}, \trialbasis^i, \testbasis^j)  \right]_{j \in \iset{\testspacena}, i \in \iset{\trialspacena}}
	&=  \sum_{T\in \grid{\Omega}} P_T^{\testspacena} \underbrace{\left[ \int_T G(\genVec{x}, \trialbasis^i, \testbasis^j) \right]_{ j\in\iset{\testspacena}_T, i\in\iset{\trialspacena}_T}}_{\eqqcolon \genLocOp} \projOp{T}{\trialspacena}.
\end{align}
It is sufficient to compute the integral over $T$ for pairings $i,j$ with
overlapping support on $T$.
$\projOp{T}{\trialspacena} : \mathbb R^{|\iset{\trialspacena}|} \rightarrow \mathbb R^{|\iset{\trialspacena}_T|}$
and $P_T^{\testspacena} : \mathbb R^{|\iset{\testspacena}_T|} \rightarrow \mathbb R^{|\iset{\testspacena}|}$
denote the selection of DoFs with support on $T$ and the corresponding mapping
back to the set of global DoFs, respectively.

\Cref{alg:locapply} implements the local matrix-vector multiplication on an arbitrary element $T$.
In practice, the integral over $T$ is transformed to a reference element $\hat{T}$ and evaluated by a quadrature rule with points $\genVec{\hat{x}}_q$ on $\hat{T}$ and weights $w_q$. The map $F : \hat{T} \rightarrow T$ facilitates the transformation of $G, \trialbasis$ and $\testbasis$ to their
reference versions $\hat{G}, \hat{\trialbasis}$ and $\hat{\testbasis}$.
In this paper we only consider the case that the map is affine, in which case its Jacobian $J_F$ is constant on each structured subdomain, i.e.,

macro-tetrahedron.
\Cref{alg:locapply} is applied inside a loop over all grid elements which is
subject of \cref{sec:LoopStrategies}.

\begin{algorithm}[t]
	\footnotesize
	\caption{Apply the local operator $\genLocOp$ on element $T$.}
	\label{alg:locapply}
	\begin{algorithmic}[1]
		\Function{LocalApply}{$T$, $\genVec{v}$, $\genVec{w}$}
			\State $\genLocOp \gets \left[ \sum_q |\text{det } J_F| w_q \hat{G}(\genVec{\hat{x}}_q, \hat{\trialbasis}^i, \hat{\testbasis}^j) \right]_{j\in\iset{\testspacena}_T, i\in\iset{\trialspacena}_T}$ \Comment{assemble local operator}
			\State $\genVec{w} \gets \genVec{w} + P_T^{\testspacena} \genLocOp \projOp{T}{\trialspacena}\genVec{v}$ \Comment{apply}
		\EndFunction
	\end{algorithmic}
\end{algorithm}

\begin{remark}
	The local matrix-vector multiplication is not necessarily split into the
	assembly of the local matrix $A_T$ and the subsequent matrix-vector
	multiplication.
	The local quadrature approach described in~\cite{Kronbichler:2012:CellBasedOA} fuses
	the two operations, such that each entry of the local result vector
	$\genLocOp \projOp{T}{\trialspacena}\genVec{v}$ is evaluated using only a single
	integral. The \fogshort does this implicitly by unrolling quadrature loops and fusing the assembly and local matrix-vector multiplication during a \ac{CSE}.
\end{remark}

\section{Optimizing Cache-Locality: Loop Strategies}
\label{sec:LoopStrategies}

On hybrid tetrahedral grids, the sum over all elements $T$ of the grid $\grid{\Omega}$ in
\cref{eq:matrixfreeop} results in an outer loop over the (unstructured)
macro-elements $\macro$ of the macro (coarsest)
grid $\macroGrid{\Omega}$ and an inner loop over the (structured) micro-elements $T_m$ which
arise from each macro due to uniform refinement.
We refer to the specific order in which the micro-elements of a single macro-element are traversed as \emph{loop strategy}.

\subsection{Sawtooth Loop Strategy}

Using the \textit{sawtooth} loop strategy, the micro-elements $\micro$ of the
current macro-element $\macro$ are iterated over in six separate loops, one for
each orientation of micro elements. Each of these \emph{element-loops} consists of a triple-nested spatial loop. \Cref{alg:sawtooth} implements the grid loop with the sawtooth loop
strategy. Selecting a micro element of a specific orientation $\orientation$
and position $\position$ in space within a macro element $\macro$ is denoted by
$\microsInMacro{\orientation}{\position}$.
$B(\orientation, \cdot)$ determines the loop bound,
which depends on the spatial orientation of the micro-element.
Specifically, we have $B(\whiteup, n) = n$, $B(\orientation, n) = n - 1$ for
$\orientation \in \{ \blueup, \bluedown, \greenup, \greendown$ \}, and
$B(\whitedown, n) = n - 2$~\cite{Kohl:2023:FundamentalDataStructures}.

\begin{algorithm}[!h]
	\footnotesize
	\caption{Sawtooth loop strategy.}
	\label{alg:sawtooth}
	\begin{algorithmic}[1]
		\Function{ElementwiseApplySawtooth}{$\genVec{v}$, $\genVec{w}$, $\ell$}
		\For{\textbf{each} $\macro \in \macroGrid{\Omega}$ } \Comment{loop macro elements}
		\For{\textbf{each} $\orientation \in \allOrientations $}  \Comment{loop orientations}
		\For{$z=0 \ldots B(\orientation, 2^\ell)$} \Comment{element loop}
		\For{$y=0 \ldots B(\orientation, 2^\ell - z)$}
		\For{$x=0 \ldots B(\orientation, 2^\ell - z - y)$}
		\State \Call{LocalApply}{$\microsInMacro{\protect \orientation}{x, y, z}$, $\genVec{v}$, $\genVec{w}$}
		\EndFor
		\EndFor
		\EndFor
		\EndFor
		\EndFor
		\EndFunction
	\end{algorithmic}
\end{algorithm}

\begin{figure}[t]
	\footnotesize
	\centering
	\vspace{-5pt}
	\subfloat{
		\resizebox{0.24\linewidth}{!}{\input{figures/interlaced_micro_cells_P2}}
	}
	\subfloat{
		\includegraphics[scale=0.06]{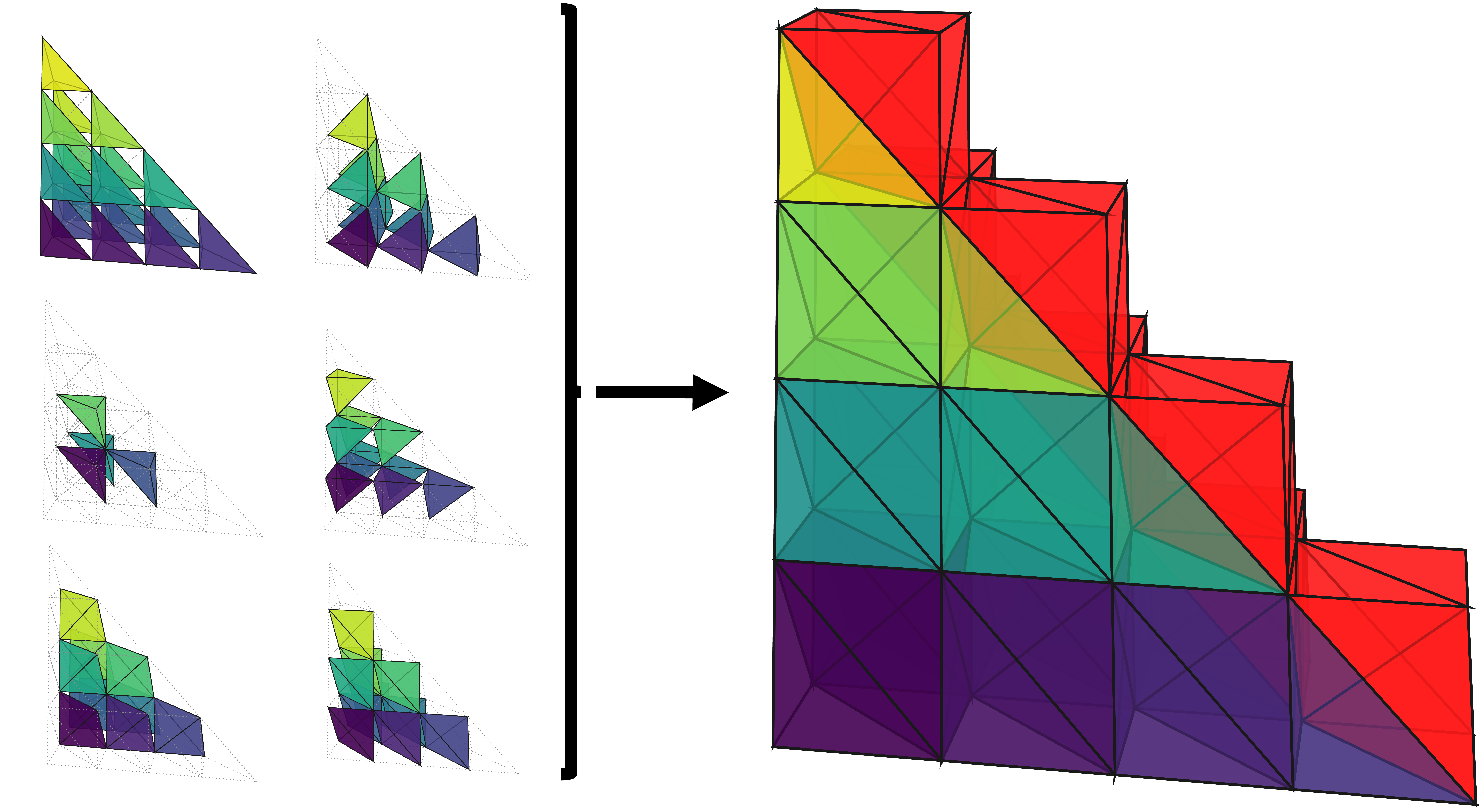}
	}
	\subfloat{
		\includegraphics[scale=0.07]{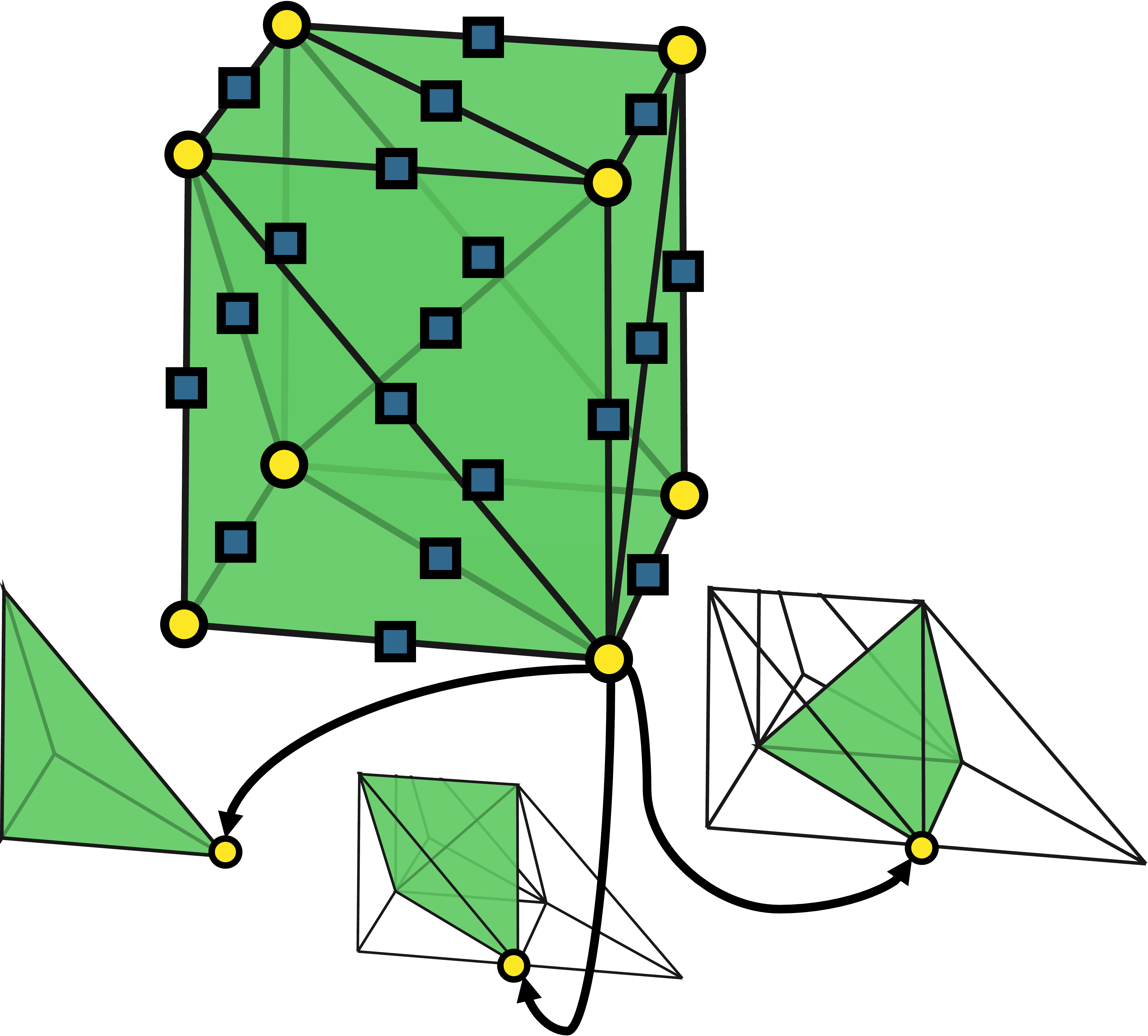}
	}
\vspace{-0.1cm}
	\caption{\textbf{Left}: Micro-elements of the
	\protect \whiteup\hspace{-0.1cm}- (purple to yellow gradient) and
	\protect \whitedown\hspace{-0.1cm}-type (in red) share vertex and edge
	DoFs. The DoFs are loaded from main memory during the
	\protect \whiteup \hspace{-0.1cm}-element loop, which then iterates the
	remaining macro-tetrahedron. The overlapping DoFs are evicted from cache
	in the process for practically relevant refinement levels. The following
	\protect \whitedown\hspace{-0.1cm}-element-loop iterates over the
	red micro-elements and has to reload the
	overlapping DoFs from main memory.
	\textbf{Middle}: The 6 element loops of
	the sawtooth loop strategy are fused together. Red micro elements lie
	outside the macro-tetrahedron and must be omitted through conditionals and
	by splitting the iteration space into complete and incomplete iterations.
	\textbf{Right}: The vertex-DoF in the lower front-right corner is part
	of multiple micro tetrahedra and will be loaded only once from main
	memory with the cubes loop strategy. }
	\label{fig:cubes}
	\vspace{-0.75cm}
\end{figure}

\cref{alg:sawtooth} is straightforward to implement
but has suboptimal memory properties for conforming discretizations. DoFs
are shared between micro-elements, therefore, DoFs are accessed
repeatedly in multiple successive element-loops of \cref{alg:sawtooth}.
After a certain element-loop loaded a specific DoF from main memory,
it will iterate the whole remaining macro-tetrahedron, such that the DoF is
likely evicted from cache.
When the following element-loop accesses the same DoF, it has to load it from
main memory again.
This leads to more main memory traffic than necessary.
\cref{fig:cubes} (left) depicts a concrete example.

\subsection{Cubes Loop Strategy}
\label{sec:CubesLoopStrategy}
The disadvantageous memory properties of the sawtooth loop strategy can be
alleviated by fusing the 6 element-loops.
The single, fused loop computes the operator application on 6 micro-elements per
iteration, which compose a cube.
If a \ac{DoF} is part of multiple micros within the same cube, as exemplified in \cref{fig:cubes} (right), all accesses to this \ac{DoF} from these micros are within a single iteration of the cubes loop.
Due to this temporal locality, the \ac{DoF} is kept in cache and must not be reloaded from main memory.
Thus, the cubes loop strategy can be considered a form of spatial blocking on tetrahedral grids.
The cubes loop strategy is depicted in \cref{fig:cubes} (middle)
and implemented in \cref{alg:cubes}.

\begin{algorithm}[!h]
	\footnotesize
	\caption{Cubes loop strategy.}
	\label{alg:cubes}
	\begin{algorithmic}
		\Function{ElementwiseApplyCubes}{$\genVec{v}$, $\genVec{w}$, $\ell$}
			\For{\textbf{each} $\macro \in \macroGrid{\Omega}$ } \Comment{loop macro elements}
			\For{$z=0\ldots 2^\ell$} \Comment{triple nested spatial loop}
			\For{$y=0\ldots 2^\ell - z$}
				\For{$x=0\ldots 2^\ell - z - y - 2$}  \Comment{complete iterations}
				\For{\textbf{each} $\orientation \in \allOrientations $}
					\State \Call{LocalApply}{$\microsInMacro{\protect \orientation}{x, y, z}$, $\genVec{v}$, $\genVec{w}$}
				\EndFor
				\EndFor
				\For{\textbf{each} $\orientation \in \{\whiteup, \greenup, \greendown, \blueup, \bluedown\}$} \Comment{incomplete iterations}
				\State \Call{LocalApply}{$\microsInMacro{\protect \orientation}{2^\ell - z - y - 1, y, z}$, $\genVec{v}$, $\genVec{w}$}
				\EndFor
				\State \Call{LocalApply}{$T_M(\protect \whiteup, 2^\ell - z - y, y, z), \genVec{v}, \genVec{w}$}
			\EndFor
			\EndFor
			\EndFor
		\EndFunction
	\end{algorithmic}
\end{algorithm}

Naively, cubes-iterations at the diagonally oriented plane of the macro tetrahedron
include micro-elements that
are not part of the macro tetrahedron (marked red in \cref{fig:cubes} (middle)).
These micros could be excluded from the iteration using conditionals.
However, to avoid conditionals in the innermost loop, the $x$-loop is cut into
three parts instead.
In all but the last two iterations, all elements of a cube lie within the macro-element.
Thus, we call these iterations \emph{complete}.
On the other hand, the remaining two iterations are \emph{incomplete}.
Neither includes the $\whitedown$-tetrahedron, and
in the last cube, only the $\whiteup$-tetrahedron is traversed.

\subsection{Memory Volume and Layer Conditions}
\label{sec:LowerandUpperBoundsforMemoryTraffic}

In the following, we evaluate cache locality of both loop strategies
and validate the assumptions made in the previous sections.
To that end, we bound the main memory volume $M$, that is how much data the considered
operators read from and write to main memory, by theoretical lower and upper bounds $M_{\text{lower}} \le M \le M_{\text{upper}}$.
If $M$ is close to $M_{\text{lower}}$, the loop strategy is efficient,
but for a naive strategy $M$ might be close to $M_{\text{upper}}$.
Determining $M_{\text{lower}}$ is straightforward: the operator has to
read and write each \ac{DoF} once from main memory in order to update
its content by a matrix-vector multiplication.
$M_{\text{upper}}$ represents the worst case: the operator reloads all associated \acp{DoF} on each micro-element
from main memory, even if they have already been accessed on previous elements.

The memory volume can be estimated more precisely through layer
conditions~\cite{Stengel:2014:StencilPerfBottlenecks, Hammer:2017:kernkraft}.
We translate the concept of layer conditions, which was originally developed for
stencil codes, to conforming, elementwise finite element operators.
To that end, we assume that a \ac{DoF} can be read from cache if and only if \emph{all} \acp{DoF} of at least one
neighboring micro-element
still reside in cache.

\begin{figure}[t]
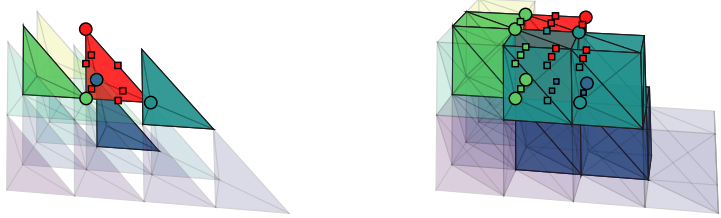

	\footnotesize
	\vspace{-5pt}
	\null\hfill
	\subfloat{
		\resizebox{0.3\linewidth}{!}{\input{figures/highlighted_lcs_sawtooth_p2}}
	}\hfill
	\subfloat{
		\resizebox{0.3\linewidth}{!}{\input{figures/highlighted_lcs_cubes_p2}}
	}\hfill\null
\vspace{-0.2cm}
	\caption{Categorization of vertex and edge DoFs on the current iteration (red) of a
	sawtooth (\textbf{left}) and cubes loop (\textbf{right}): light green
	DoFs are always cached from the
	previous iteration (light green elements), red DoFs are never cached because they lie ahead of the iteration order, teal and blue DoFs are cached depending on the presence of data from the corresponding neighbor elements in cache.}
	\label{fig:lcs}
	\vspace{-20pt}
\end{figure}

\Cref{fig:lcs} (left) illustrates this for the $\whiteup$-part of the
sawtooth loop-strategy for vertex and edge \acp{DoF} of a $\quadPoly$ discretization. 
The \acp{DoF} of the current micro-tetrahedron (red) are subdivided into
different groups: The light green \acp{DoF} will always be in cache
because they have been accessed in the previous iteration.
The red \acp{DoF} will never be in cache because they lie ahead in the iteration
order  (assuming a single operator application).
For teal and blue \acp{DoF} it depends on layer conditions, i.e., whether the
neighbor element with the corresponding color has been evicted.
Put differently, whether the \textit{tail} of the iteration containing those elements is still in cache.
In the cubes loop \cref{fig:lcs} (right), a whole front of \acp{DoF} can be read from cache from the last cube iteration (light green). They make up $\nicefrac{1}{3}$ of all \ac{DoF} accesses. With the sawtooth loop strategy, this fraction is just $\nicefrac{1}{10}$.

In practice, we compute an estimate for the main memory volume based on layer conditions $M_{\text{lc}, s}, \ s \in \{\text{sawtooth}, \text{cubes} \}$ as follows.
We know the memory volumes $M_s^{\text{red}}$, $M_s^{\text{teal}}$ and $M_s^{\text{blue}}$ associated to the respective \ac{DoF} accesses per iteration of a loop strategy $s$. Furthermore, we can compute the size of the tails $M_{i, s}^\text{tail, teal}$ and $M_{i, s}^\text{tail, blue}$ between the current element in iteration $i$ and the elements that have to reside in cache to provide an overlapping \ac{DoF} (teal or blue colored). 
Then, we compare the computed tail size with the actual size of the cache on the used machine, e.g.~the L3 cache. Doing this for each micro-element on each macro-element, we obtain an estimate for the memory volume an operator requires during its matrix-free application:

\begin{align}
	\label{eq:lc-est}
	 M_{\text{lc}, s} = n_\text{macros} \cdot \sum_{i} M_s^{ \text{red}} + \begin{cases}
	 	M_s^{ \text{teal}} \text{ if } M_{i, s}^\text{tail, teal} \ge M_\text{L3} \\
	 	M_s^{ \text{blue}} + M_s^{ \text{teal}} \text{ if } M_{i, s}^\text{tail, blue} \ge M_\text{L3} \\
	 \end{cases} 
\end{align}
with $n_\text{macros}$ being the number of macro tetrahedra on the MPI process.
If the tail for a certain neighboring element, e.g.~the teal colored ones in \cref{fig:lcs}, is larger than the cache size, it must have been evicted from cache and reloaded from main memory on access. This is a simplified model that neglects certain aspects of the caching behavior, but yields good results in practice.

\subsection{Memory Study}
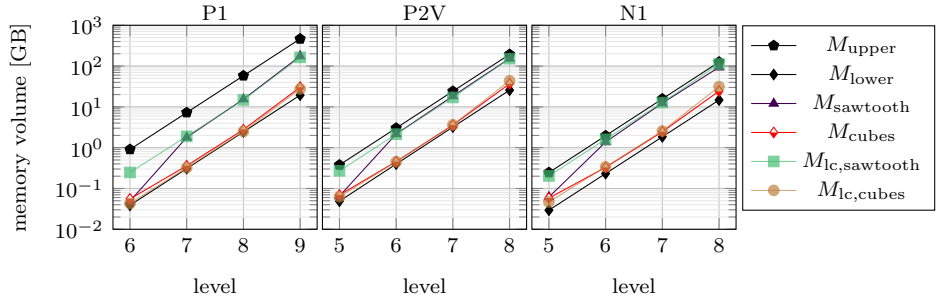
\begin{figure}
\begin{tikzpicture}
	\sisetup{exponent-mode=scientific, round-mode=places}
	\begin{axis}[ width  = 0.33\textwidth,
		name=p1,
		, ymin=0.01, ymax=1000,
		, height = 0.33\textwidth
		, xlabel = {level}
		, ylabel = {memory volume [GB]}
		, title  = {P1}
		, title style = {at={(0.5,0.9)}}
		, xtick distance   = 1
		, ytick distance   = 10^1
		, ymode            = log
		, grid             = both
		, legend pos       = north east
		, mark options = {scale = 1}
		]
		\addplot[color=black,mark=diamond*] table[y = lw] {data/mem_vols_p1.dat};
	\addplot[color=black,mark=pentagon*] table[y = up] {data/mem_vols_p1.dat};
	\addplot[color of colormap={0},mark=triangle*] table[y = st] {data/mem_vols_p1.dat};
	\addplot[color=red,mark=halfdiamond*] table[y = cb] {data/mem_vols_p1.dat};
	\addplot[color of colormap={700},mark=square*, opacity=0.7] table[y = lcst] {data/mem_vols_p1.dat};	
	\addplot[color=brown,mark=*, opacity=0.7] table[y = lccb] {data/mem_vols_p1.dat};
		\end{axis}
	\hspace{2pt}
	\begin{axis}[ width  = 0.33\textwidth,
		name=p2v, at={(p1.south east)},
		yticklabel=\empty,
		, ymin=0.01, ymax=1000,
		, height = 0.33\textwidth
		, xlabel = {level}
		, title  = {P2V}
		, title style = {at={(0.5,0.9)}}
		, xtick distance   = 1
		, ytick distance   = 10^1
		, ymode            = log
		, grid             = both
		, legend pos       = north east
		, mark options = {scale = 1}
		]
		\addplot[color=black,mark=diamond*] table[y = lw] {data/mem_vols_p2v.dat};
		\addplot[color=black,mark=pentagon*] table[y = up] {data/mem_vols_p2v.dat};
		\addplot[color of colormap={0},mark=triangle*] table[y = st] {data/mem_vols_p2v.dat};
		\addplot[color=red,mark=halfdiamond*] table[y = cb] {data/mem_vols_p2v.dat};
		\addplot[color of colormap={700},mark=square*, opacity=0.7] table[y = lcst] {data/mem_vols_p2v.dat};	
		\addplot[color=brown,mark=*, opacity=0.7] table[y = lccb] {data/mem_vols_p2v.dat};
	\end{axis}
    \hspace{2pt}
    \begin{axis}[ width  = 0.33\textwidth,
	name=n1, at={(p2v.south east)},
	yticklabel=\empty,
	, ymin=0.01, ymax=1000,
	, height = 0.33\textwidth
	, xlabel = {level}
	, title  = {N1}
	, title style = {at={(0.5,0.9)}}
	, xtick distance   = 1
	, ytick distance   = 10^1
	, ymode            = log
	, grid             = both
	,  legend pos=outer north east
	, mark options = {scale = 1}
	]
	\addplot[color=black,mark=pentagon*] table[y = up] {data/mem_vols_n1.dat};
	\addplot[color=black,mark=diamond*] table[y = lw] {data/mem_vols_n1.dat};
	\addplot[color of colormap={0},mark=triangle*] table[y = st] {data/mem_vols_n1.dat};
	\addplot[color=red,mark=halfdiamond*] table[y = cb] {data/mem_vols_n1.dat};
	\addplot[color of colormap={700},mark=square*, opacity=0.7] table[y = lcst] {data/mem_vols_n1.dat};	
	\addplot[color=brown,mark=*, opacity=0.7] table[y = lccb] {data/mem_vols_n1.dat};
	    \legend{
		$M_{\text{upper}}$,
		$M_{\text{lower}}$,
		$M_{\text{sawtooth}}$,
		$M_{\text{cubes}}$,
		$M_{\text{lc,sawtooth}}$,
		$M_{\text{lc,cubes}}$,
	}
\end{axis}
\end{tikzpicture}
\vspace{-0.2cm}
\caption{Corridor plot of main memory volume against refinement levels,
comparing the measurements with the analytical bounds
	and layer condition estimates.
 }
\label{fig:corridorplot}
\vspace{-0.75cm}
\end{figure}

\cref{fig:corridorplot} compares the measured main memory volume for the two loop strategies $M_{\text{sawtooth}}$, $M_{\text{cubes}}$, obtained by \likwid's \cite{Wellein:2010:Likwid} hardware performance counters, with the bounds $M_{\text{upper}}$, $M_{\text{lower}}$ and layer condition estimates $M_{\text{lc, sawtooth}}$, $M_{\text{lc, cubes}}$ based on \cref{eq:lc-est}. 
We measure \cref{eq:p1diff}, a constant diffusion operator ($-\Delta$) discretized by linear Lagrangian elements, \cref{eq:p2divkgrad}, a variable-coefficient diffusion operator ($-\nabla \cdot (k(\mathbf{x})\, \nabla\,)$) discretized by quadratic Lagrangian elements and \cref{eq:n1e1curlcurlplusmass}, a curl-curl operator ($\alpha(\mathbf{x})\, \curl \curl 
+ \beta(\mathbf{x}) $) discretized by N\'{e}d\'{e}lec elements (more details
on the operators in \cref{sec:PerfEng}).

The upper and lower bounds define a
corridor in which all measurements lie. The cubes loop strategy consistently comes close to
the lower bound, showing high cache locality and only few, necessary accesses to main memory.

In terms of cache locality, it clearly outperforms the sawtooth loop strategy
which consistently places close to the upper bound.
On low refinement levels both strategies achieve very low memory volumes close
to the lower bound. For these cases the data of a complete
macro-tetrahedron fits into the L3-cache such that no special strategy is required to
improve caching.

Furthermore, the estimates incorporating layer conditions closely match the
measurements for both loop strategies. On level 9 for \cref{eq:p1diff} and 8
for \cref{eq:p2divkgrad} and \cref{eq:n1e1curlcurlplusmass}, the stronger (blue) layer
condition breaks, and we see an upward kink in the measured memory volume for
both loop strategies (less visible for sawtooth due the breaking of layer conditions only contributing a single \ac{DoF}). Apparently, our estimates precisely capture the breaking of layer conditions.

Complementary to these pure memory focused observations,
\cref{sec:PerfEng} analyses the impact of the cubes loop strategy on
kernel performance.

\section{Optimizing Computations}
\label{sec:genOpts}
In the following, we present the optimizations the code generator applies to speed-up computation
and reduce redundant calculations, and how they are realized as \ac{AST} transformations during generation time.
In \cref{sec:PerfEng} we will apply them to
three different bilinear forms.

\subsection{Automatic Identification of Loop Invariants}

The evaluation of a bilinear form over elements of a structured grid may
contain large fractions of redundant computation. For example, due to the
translation invariance of micro-elements of the same
orientation~\cite{Kohl:2023:FundamentalDataStructures},
their Jacobians are identical and therefore loop-invariant.
It is common that in this and similar cases, the computation is moved outside
of the loop, see~\cite{Olgaard:2010:OptimQuadRepr} for
more examples.

Such invariants can be identified automatically by the \fogshort \ via traversing
the \ac{AST} and checking statements in loop bodies for a dependency on loop
counters.
If there is no dependency, the statement is moved in front of the loop,
eliminating the redundancy and reducing the number of \acp{FLOP}. The \fogshort thereby not only targets the local operator
application by drawing invariants out of the loop over quadrature points, but
also recognizes computation that is spatially-invariant.

\subsection{Inter-element Vectorization}

Automatic vectorization of the grid loop of the matrix-free operator
application is a significant challenge for the backend \cpp compiler.
This is due to the irregular iteration space on tetrahedra and a large loop body
possibly containing another loop over quadrature points.
On the other hand, handwriting vectorized versions of the assembly for different
vector widths, instruction sets, architectures, weak forms and discrete function
spaces is extremely tedious.

The \fogshort \ automatically vectorizes the
matrix-free operator application during generation time.
It cuts the $x$-direction loop into a vectorized and a remainder loop.
All AST-nodes in the body of the former are replaced by vector
instructions and the loop counter is modified to run over patches of multiple
elements.
The result is an operator that computes the local assembly on multiple
micro-elements simultaneously, an inter-element vectorization similar
to~\cite{Sun:2019:Vect}.

\subsection{Integration}

The type of quadrature rule used to compute the integrals in the local
assembly significantly affects the number of \acp{FLOP} an operator executes. A quadrature
rule that achieves exact integration with the fewest points is an obvious
choice. However, according to~\cite{Ciarlet:2002:FE}, for second-order elliptic variational problems with polynomial degree $q$ in the FE spaces, a quadrature rule exact up to polynomials of degree $2q - 2$ is sufficient to achieve the expected convergence rate. This is called \textit{under-integration}~\cite{Kirby:2006:FFC}.
Alternatively, the \fogshort can integrate the local matrix analytically, providing the option of a quadrature-free kernel.

\subsection{Common Subexpression Elimination}
Two types of \ac{CSE} are available in the \fogshort: a traditional, tree-based \ac{CSE} and a polynomial \ac{CSE} following \cite{Hosangadi:2006:polyCSE} which may be more effective for expressions that resemble polynomials. Examining their effectivity on matrix-free \ac{FE} operators is out of scope of this paper. 

\subsection{Quadrature Loops}
The generator can either unroll the loops over quadrature points in \cref{alg:locapply} or generate them in loop form. In the latter case, the quadrature loops for each entry of the local matrix are fused into a single loop to improve the \ac{CSE} applied to the loop body. Unrolling the quadrature loops offers a wider space of expressions the \ac{CSE} can eliminate on, which leads to more effective elimination and fewer \acp{FLOP}. However, it can bloat the number of statements in the kernel, leading to L2 cache problems. We investigate this in \cref{sec:InstBoundedness}. 

\subsection{Tabulation of Factors of the Weak Form}

Another common technique orthogonal to precomputing whole local
element matrices is to \emph{tabulate}, that is to precompute factors of the weak form, e.g., the shape functions and their gradients, store them in table-like structures and access them in the kernel. This is e.g., done by the \basix \  package within the \fenics \ project \cite{Alnaes:2015:FEniCSProjectVersion}.

Consider the entries of the local element matrix that arise from the
discretization of the operator $-\nabla \cdot (k(\mathbf{x})\, \nabla\,)$:
\begin{equation}\label{eq:tabulation-example}
	\begin{aligned}
		\genLocMicroOp =
		\locMat{\sum_{q} k(\genVec{\hat{x}}_q) \underbrace{w_q|\text{det }
				J_F|  (J_F^{-T}\hat{\nabla} \hat{\basis}^i(\genVec{\hat{x}}_q) \cdot J_F^{-T}\hat{\nabla} \hat{\basis}^j(\genVec{\hat{x}}_q))}_{\star}}{\iset{\quadPoly}_{T_m}}{\iset{\quadPoly}_{T_m}}.
	\end{aligned}
\end{equation}
The gradients in \cref{eq:tabulation-example} are computed solely at quadrature points on the reference element $\hat{T}$ and independent of the micro element $T_m$. They can be tabulated for each  shape function $\trialbasis^i$ and quadrature point $\genVec{\hat{x}}_q$. The table has $10 \cdot n_q$ vector-valued entries for 10 shape functions in $\iset{\quadPoly}_{\micro}$ and $n_q$ quadrature points. It is accessed in the loop over micro elements and the values used to compute the local assembly.

On hybrid tetrahedral grids, the important property that micro elements of a certain orientation $\orientation$ are translation-invariant \cite{Kohl:2023:FundamentalDataStructures} leads to the Jacobian of the affine mapping $J_F$ being identical
for all micro elements of that orientation. \textit{Therefore, on hybrid
tetrahedral grids, not only the gradient in
\cref{eq:tabulation-example} can be tabulated, but the complete factor $\star$.} Then, we tabulate for each pairing of shape function $(\trialbasis^i, \trialbasis^j)$, quadrature point and orientation of micro-elements, yielding a table with $6 \cdot 100 \cdot n_q$ scalar entries, when neglecting symmetry.

\begin{remark}
	In case of an additional curvilinear transformation occurring on certain meshes, another, spatially dependent Jacobian enters the weak form and we have to tabulate multiple factors.
\end{remark}

\subsection{Symmetry}
A symmetric variational form implies a symmetric local matrix. For such forms only half of the off-diagonal entries have to be computed, an optimization that reduces the \acp{FLOP} required for local assembly. 

In moderately sized \acp{AST}, the CSE is able to detect such symmetries itself without any additional effort.
However, for large \acp{AST} with many nodes, e.g., stemming from high degree quadrature rules or complicated integrands, the CSE sometimes fails to detect symmetric parts of the local matrix. Fortunately, the optimization is straightforward to implemented as an AST transformation: the entries of the local matrix are available as symbolic expressions and the symmetric entries can be replaced with accesses to their counterparts.
\subsection{Precomputation of Local Element Matrices}
\label{sec:PrecomputationLocalElementMatrices}
A common technique to speed up the matrix-vector multiplication of an \ac{FE} operator is to
compute the local element matrix for each element and store it a priori~\cite{Carey:1988:elembyelem,Carey:1986:CANM}.
Each time the operator is applied, the stored local matrix of the current
element in the iteration is loaded from memory and applied to the local DoFs.
This is especially advantageous for complicated variational forms and high-order
quadrature rules, because all the implied computation can be shifted to a setup phase.
However, the major drawback of the approach is the massively increased demand in
main memory volume that makes it only feasible for relatively low refinement levels.
\emph{Furthermore, an operator using precomputation essentially only runs as fast as
	the local matrices can be loaded from memory and is thereby deeply memory-bound.}
We include this here only for reference, as this approach is not matrix-free.

\section{Performance Analysis}
\label{sec:PerfEng}

We evaluate the optimizations from \cref{sec:LoopStrategies,sec:genOpts}
implemented in the HOG by generating a range of operators for three different bilinear forms and
different finite element spaces.
Particularly, we generate the matrix-free matrix-vector multiplication optimized for
maximum throughput measured in updated DoF/s on a single node of the
architecture presented in \cref{subsec:Fritz}.

We will make use of the roofline performance
model~\cite{Williams:2009:Roofline} which serves as a simple, yet effective tool to
understand the measured performance, observe the impact of optimizations, identify bottlenecks and guide the
optimization process. The peak performance, maximum memory bandwidth and other roofline-constants are measured using \likwid \cite{Wellein:2010:Likwid}.
The general approach is to explore the search space that is given by all
combinations of optimizations and determine the fastest operator for each
weak form, for which we lay out the optimization path taken to obtain it.

We emphasize three optimizations here, because the \fogshort applies them to all weak forms in this article and they can be considered as a standard approach to optimize matrix-free, elementwise \ac{FE} operators (on hybrid tetrahedral grids):
\begin{enumerate}
	\item Speed-up arithmetic by vectorizing across elements and executing
	arithmetic in parallel on multiple vector lanes. This shows as a
	performance boost and straight upward shift in the roofline model plot.

	\item Eliminate redundant computation by identifying loop invariants and moving them ahead of the loop. This reduces the arithmetic intensity and shifts operators in direction of the memory-bound region of the roofline.
	\item Alleviate the memory-boundedness induced during the previous step with
	the cubes loop strategy. Improving the cache-locality increases the arithmetic intensity and manifests as a shift in the direction of the compute-bound region of the roofline. A performance boost is caused by the reduction of the main memory volume in memory-bound operators.
\end{enumerate}

Each partially optimized variant of an operator is labeled by a prefix for
each of the test cases (\cref{eq:p1diff}, \cref{eq:p2divkgrad},
or \cref{eq:n1e1curlcurlplusmass}) that are described at the start of the
following sections.
They are separated by an '\_' from a list of letters encoding the applied
optimizations, which are given in \cref{tab:Opts}. For instance, P1\_SV represents matrix-free operator for the diffusion form with optimizations symmetry (S) and vectorization (V).

\subsection{Test Case and Machine}

\label{subsec:Fritz}
A single matrix-free application in double precision serves as the main test application.
We do not consider the communication prior to and after the operator
application, as this is out-of-scope for this work but will be considered in future
publications.

The experiments are conducted on refinement level 7.
Generally, we want to use as few macro-elements as possible while still accurately capturing the domain. This way, we can use many refinements, yielding largely ordered structure and potential for high performance.
Our previous applications indicate that the refinement level typically lies
in the range from 5 to 8.
In order to distribute at least one macro-primitive to each process,
the coarse grid should also have at least as many macro-primitives as there are
processes.
For this paper, we choose a simple cuboid geometry made
up of 36 macro-tetrahedra for the coarsest grid $\macroGrid{\Omega}$.

\begin{table}[t]
	\footnotesize
	\begin{minipage}[b]{0.43\linewidth}
		\centering
			\begin{tabular}{ll}
			\toprule
		 optimization & short \\
			\midrule
			 symmetry  & S\\
			 inter-element vectorization & V \\
			 loop invariants & I \\
			 cubes loop strategy & C\\
			 under-integration & U \\
			 fused quadrature loops & fQ \\
			 tabulation & T\\
			 precomputation & P\\
			\bottomrule
		\end{tabular}
		\caption{Range of optimizations from \cref{sec:genOpts} and their abbreviations.
		}
		\label{tab:Opts}
	\end{minipage}
	\hspace{0.5cm}
	\begin{minipage}[b]{0.5\linewidth}
		\vspace{-2cm}
		\centering
		\begin{tabular}{lrr}
			\toprule
			cores & 36 & (per socket) \\
			L1data cache size &  48 KB & (per core)\\
			L1inst cache size &  32 KB & (per core)\\
			L2 cache size &  1.25 MB & (per core) \\
			L3 cache size &  54 MB & (shared)\\
			clock speed   & 2.6 GHz & (fixed) \\
			\bottomrule
		\end{tabular}
		\caption{Relevant technical details for the Intel(R) Xeon(R) Platinum 8360Y CPUs of type IceLake SP.}
		\label{tab:fritzStats}
	\end{minipage}
\vspace{-0.75cm}
\end{table}

The macro-primitives are distributed to the 36 cores of a single socket of
the Fritz supercomputer at the Erlangen National High Performance Computing Center
NHR@FAU~\cite{NHR:2022:Fritzonline}.
\Cref{tab:fritzStats} summarizes relevant information regarding the cache hierarchy, clock speed and number of cores of the CPU type built into the Fritz supercomputer. Measurements of roofline constants and the test application are conducted with likwid-bench and likwid-mpirun from the \likwid performance-measurement tools \cite{Wellein:2010:Likwid}. The clock frequency is fixed to 2.6 GHz by the slurm environment on Fritz and processes are pinned to processors by \likwid pinning masks. The applications are compiled with Intel ICX 2021.4 and the \texttt{-march=native -O3} flags. 

The reference implementations of the elementwise operators, on which the
optimizations build, 
are generated by the \fogshort without optimizations and do not yet take the regular grid structure into account.
Furthermore, apart from running
\sympy's \ac{CSE} during generation of the local assembly, they fully rely on the
backend \cpp compiler for optimizations like vectorization.

\subsection{Constant Diffusion}
The simplest test case benchmarks a constant diffusion operator $-\Delta$
discretized by linear continuous Lagrangian elements, i.e., $\trialspacena = \testspacena = \linPoly(\microGrid{\Omega})$. Despite its simplicity, its performance characteristics are representative for compute-sparse, low-order operators, such as
weak divergence, gradient, or stabilization terms for the mixed $\linPoly-\linPoly$ approximation of the Stokes equation
used in \cite{Hughes:1986:stabilizedStokes}.
The local operator is given by:
	\begin{equation}
		\label{eq:p1diff}
		\genLocMicroOp =\locMat{\intMicroElem \nabla \trialbasis^i \cdot \nabla \testbasis^j}{\iset{\linPoly}_{T_m}}{\iset{\linPoly}_{T_m}} \tag{\textbf{P1}}
	\end{equation}
with $\iset{\linPoly}_{T_m}$ the index set of
linear basis functions with support on micro element $T_m$. The linear spaces yield $1.3 \times 10^8$ \ac{DoF}s on the test cube at level 7.

\begin{figure}
  \subfloat{
    \begin{tikzpicture}
    	  \begin{loglogaxis}
        [ roofline
        , title = {optimization search}
        , xmax=40
        ]
        \drawroofline
		\addplot[color of colormap={100},mark=halfsquare right*, opacity=0.5] coordinates {(7.392377698700969, 286.0341973)}; 
		\addplot[color of colormap={200},mark=halfsquare left*, opacity=0.5] coordinates {(8.85184411422856, 279.8733415)}; 
		\addplot[color of colormap={300},mark=triangle*] coordinates {(3.6861797271348333, 256.7792057)}; 
		\addplot[color of colormap={400},mark=diamond*, opacity=0.5] coordinates {(7.291691823930124, 307.9367244)}; 

        \addplot[color of colormap={700},mark=halfdiamond*] coordinates {(2.015247239571319, 237.1467412)}; 
        \addplot[color of colormap={600},mark=halfcircle*] coordinates {(1.7091938050444404, 211.92342000000002)}; 
       
        \addplot[only marks,color of colormap={800},mark=*] coordinates {(1.4960887736329989, 210.54530499999998)}; 
        \label{pgf:P1_SVIfQ} 
        \addplot[color of colormap={0},mark=halfsquare*, opacity=0.5] coordinates {(7.439998888832885, 282.3426546)}; 
        
        \end{loglogaxis}
    \end{tikzpicture}
  }
  \subfloat{
    \begin{tikzpicture}
    	   \begin{loglogaxis}
        [ roofline
        , title = {optimization path}
        , xmax=40
        ]
        \drawroofline

        \addplot+[only marks, color of colormap={800},mark=oplus] coordinates {(6.134878049369867, 177.9694276)}; \label{pgf:P1_S} 
        \addplot+[only marks, color of colormap={400},mark=square*] coordinates {(5.54123820203525, 569.2239625000001)}; \label{pgf:P1_SV} 
        \addplot+[only marks, color of colormap={700},mark=halfcircle*] coordinates {(2.1551696280280446, 237.1467412)}; \label{pgf:P1_SVI} 
        \addplot+[only marks, color of colormap={400},mark=diamond*, opacity=0.5] coordinates {(7.291691823930124, 307.9367244)}; \label{pgf:P1_SVIC} 
		 \addplot+[only marks, color of colormap={0},mark=halfsquare*, opacity=0.5] coordinates {(7.439998888832885, 282.3426546)}; \label{pgf:P1_SVICfQ} 

        \draw[a, bend right=80, looseness=2] (axis cs:6.134878049369867, 177.9694276) to node[n]{V} (axis cs:5.54123820203525, 569.2239625);
        \draw[a]                             (axis cs:5.54123820203525, 569.2239625)  -- node[n]{I} (axis cs:2.015247239571319, 237.1467412);
        \draw[a]                             (axis cs:2.015247239571319, 237.1467412) -- node[n]{C} (axis cs:7.291691823930124, 307.9367244);
      
      \end{loglogaxis}
    \end{tikzpicture}
  }
\vspace{-0.3cm}
  \subfloat{
    \begin{tikzpicture}
    
      \begin{axis}
        [ performance
        , symbolic y coords = {P1_SVICfQ, P1_SVICT, P1_SVICfQT, P1_SVIT, P1_SVIC, P1_SVI, P1_SVIfQT, P1_SVIfQ}
        , ytick = {P1_SVICfQ, P1_SVICT, P1_SVICfQT, P1_SVIT, P1_SVIC, P1_SVI, P1_SVIfQT, P1_SVIfQ}
        , yticklabels = {P1\_SVICfQ, P1\_SVICT, P1\_SVICfQT, P1\_SVIT, P1\_SVIC, P1\_SVI, P1\_SVIfQT, P1\_SVIfQ}
        ]
        \addplot[color of colormap={0},fill=.!30,mark=halfsquare*] coordinates {(1441.7549999999999,P1_SVICfQ)}; 
        \addplot[color of colormap={100},fill=.!30,mark=halfsquare right*] coordinates {(1291.7264583333335,P1_SVICT)}; 
        \addplot[color of colormap={200},fill=.!30,mark=halfsquare left*] coordinates {(1393.7460674157305,P1_SVICfQT)}; 
        \addplot[color of colormap={300},fill=.!30,mark=triangle*] coordinates {(1107.8785714285714,P1_SVIT)}; 
        \addplot[color of colormap={400},fill=.!30,mark=diamond*] coordinates {(1319.1549999999997,P1_SVIC)}; 
        \addplot[color of colormap={700},fill=.!30,mark=halfdiamond*] coordinates {(761.4601226993865,P1_SVIfQT)}; 
          \addplot[color of colormap={800},fill=.!30,mark=*] coordinates {(1041.9,P1_SVIfQ)}; 
        
        \addplot[color of colormap={600},fill=.!30,mark=halfcircle*] coordinates {(947.5557251908396,P1_SVI)}; 
      \end{axis}
    \end{tikzpicture}
  }
  \subfloat{
    \sisetup{round-mode=figures}
    \begin{tikzpicture}
    	
    	\label{fig:p1histmdofs}
      \begin{axis}
        [ performance
        , symbolic y coords = {P1_SVICfQ, P1_SVIC, P1_SVI, P1_SV, P1_S}
        , ytick = {P1_S, P1_SV, P1_SVI, P1_SVIC, P1_SVICfQ}
        , yticklabels = {P1\_S, P1\_SV, P1\_SVI, P1\_SVIC, P1\_SVICfQ}
        ]
        \addplot[color of colormap={800},fill=.!30,mark=oplus] coordinates {(197.4691082802548,P1_S)}; 
        \draw[a, bend left] (axis cs:197.4691082802548,P1_S) to node[n, anchor=south west]{$\num{3.549640530546591}\times$} (axis cs:700.9443502824859,P1_SV);
        \addplot[color of colormap={400},fill=.!30,mark=square*] coordinates {(700.9443502824859,P1_SV)}; 
        \draw[a, bend left] (axis cs:700.9443502824859,P1_SV) to node[n, anchor=south west]{$\num{1.351827323822451}\times$} (axis cs:947.5557251908396,P1_SVI);
        \addplot[color of colormap={600},fill=.!30,mark=halfcircle*] coordinates {(947.5557251908396,P1_SVI)}; 
        \draw[a, bend left] (axis cs:947.5557251908396,P1_SVI) to node[n, anchor=south west]{$\num{1.392166143826865}\times$} (axis cs:1319.1549999999997,P1_SVIC);
        \addplot[color of colormap={400},fill=.!30,mark=diamond*] coordinates {(1319.1549999999997,P1_SVIC)}; %
         \draw[a, bend left] (axis cs:1319.15499999,P1_SVIC) to node[n, anchor=south west]{$\num{1.1}\times$} (axis cs:1441.754999999999,P1_SVICfQ);
        
        \addplot[color of colormap={0},fill=.!30,mark=halfsquare*] coordinates {(1441.7549999999999,P1_SVICfQ)}; 
      \end{axis}
    \end{tikzpicture}
  }
   
  \vspace{-0.2cm}
  \caption{\textbf{Left column}: Roofline (top) and MDoF/s (bottom) for the set of
  fastest operators of the form \cref{eq:p1diff}.
  \textbf{Right column}: Roofline (top) and MDoF/s (bottom) with speed-ups for the
  optimization path to \inst{P1_SVICfQ}. The accumulated speed-up is \num{7.3}$\times$. 
  \vspace{-0.3cm}
}
  \label{fig:p1roof}
  \vspace{-20pt}
\end{figure}
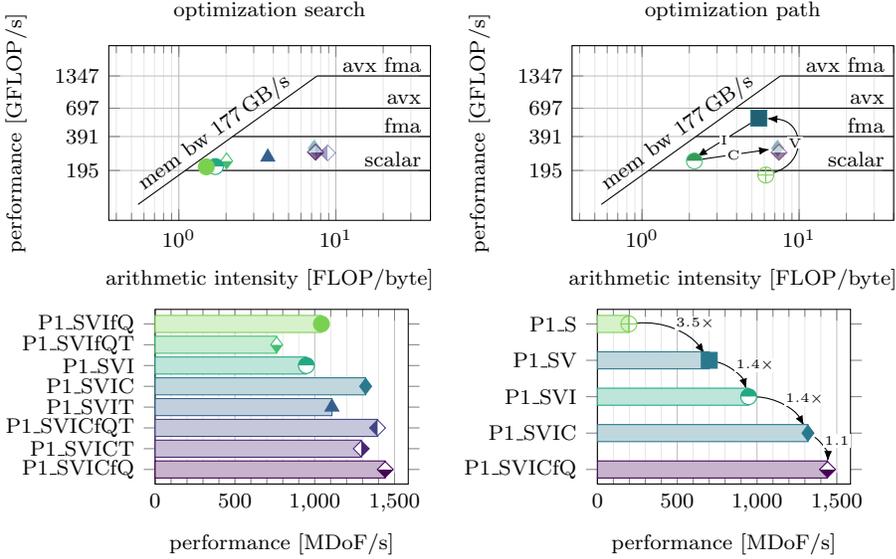

The operators using the most effective sets of optimizations with respect to \mbox{MDoF/s} are
plotted on the left column of \cref{fig:p1roof}. Remarkably, the four fastest operators all utilize the cubes loop strategy (C). Given the inherent compute-sparsity associated with low polynomial degree,
the enhancement of cache-locality becomes imperative to overcome the constraints imposed
by memory bandwidth limitations. These operators achieve a throughput of up to 1.5
GDoF/s but reach only \qty{23}{\%} of the AVX
FMA peak performance. 

The optimization path to obtain \inst{P1_SVICfQ}, the fastest operator
on the current machine is presented on the right column of \cref{fig:p1roof}. 
We start with \inst{P1_S}, an operator restricted by the scalar instruction roof.
Employing our standard optimization of inter-element AVX vectorization (\inst{P1_SV}),
the code is no more limited to the hardware roof prescribed by scalar instructions, 
and the performance shoots vertically up, reaching close to the AVX roofs.
To obtain \inst{P1_SVI}, the \fogshort identifies the whole integrand, 
that is the constant gradients of test and trial basis and the Jacobian of the pull-back, as loop invariant and moves it ahead of the loop.
This drastically decreases the overall computations while keeping the memory traffic almost constant.
The result is a corresponding decrease in the arithmetic intensity, rendering \inst{P1_SVI} memory-bound.
The performance drops in terms of FLOP/s from close to \qty{50}{\%} to \qty{23}{\%} peak performance.
However, the performance in DoF/s improves (and consequently runtime decreases) by a factor of 1.4$\times$ due to a reduction in the overall \acp{FLOP}.
Finally, we enhance the cache-locality of the memory-bound \inst{P1_SVI} by the
cubes loop strategy, recovering its arithmetic intensity and placing it close to
the machine balance. 
The final operator \inst{P1_SVICfQ} achieves an accumulated speed-up of $\sim$ 7.3$\times$ compared to the scalar version.

\begin{remark} 
Contrary to the upcoming, more compute-intensive \cref{eq:n1e1curlcurlplusmass} and \cref{eq:p2divkgrad} operators, \cref{eq:p1diff} operators are still memory-bound or at the machine-balance even after applying the cubes loop strategy. Relocating loop invariants is exceptionally effective here, removing almost all computations from the loop and essentially reducing the operator to an elementwise copy-benchmark on tetrahedral grids. Thereby, the operators are prone to bottlenecks in the data path. Although the optimization leads to speed-up, it causes lower FLOP/s performance.
This exemplifies the crucial difference between GDoF/s and FLOP/s: the latter is not always meaningful to measure performance improvements in presence of an optimization that changes the amount of \acp{FLOP}.
\end{remark}

\subsection{Curl-Curl and Mass}
\label{sec:CCPM}
Next, we evaluate a linear combination of the curl-curl and
mass operator~\cite{Hipmair:2009:MGHcurl}:
$\alpha(\genVec{x}) \ \curl \curl  + \beta(\genVec{x}) $.
Test and trial space are \textit{N\'{e}d\'{e}lec} elements of first order and first kind~\cite{Nedelec:1980:NedelecElems}.
We assume that $\alpha(\bm{x}), \beta(\bm{x}) \in \coeffspacena$, a discrete FE space
used to approximate the coefficient function and that setting $\coeffspacena = \linPoly(\microGrid{\Omega})$ yields a sufficient
approximation for smooth coefficients.
In the N\'{e}d\'{e}lec space $\nedelec(\microGrid{\Omega})$, \ac{DoF}s
 are associated with 
the edges of the element, yielding \num{e9} \acp{DoF} on the test cube. The corresponding basis functions are vector-valued
and ensure continuity across elements in tangential direction~\cite{Nedelec:1980:NedelecElems}.

The associated local operator is defined as
\begin{align}
	\label{eq:n1e1curlcurlplusmass}
	\genLocMicroOp =\locMat{\intMicroElem \alpha(\bm{x}) (\curl \bm{\trialbasis}
		^i \cdot  \curl \bm{\testbasis}^j) +  \beta(\bm{x}) (\bm{\trialbasis}^i \cdot  \bm{\testbasis}^j)}{\iset{\nedelec}_{T_m}}{\iset{\nedelec}_{T_m}}.
	\tag{\textbf{N1}}
\end{align}
\begin{figure}
  \subfloat{
    \begin{tikzpicture}
    	\label{fig:n1e1bestroof}
      \begin{loglogaxis}
        [ roofline
        , title = {optimization search}
        , xmax=40
        ]
        \drawroofline[avx=false, fma=false]
	   \addplot[color of colormap={100},mark=halfsquare*] coordinates {(2.806990483900432, 423.3016529)}; 
    \addplot[color of colormap={0},mark=halfsquare right*, opacity=0.5] coordinates {(3.463750611391227, 524.0131046)}; 
   \addplot[color of colormap={200},mark=halfsquare left*] coordinates {(7.033998539094143, 821.912862)}; 
     \addplot[color of colormap={300},mark=triangle*] coordinates {(29.094540333700465, 728.9789536000001)}; 
 		\addplot[color of colormap={400},mark=diamond*, opacity=0.5] coordinates {(17.371753637389197, 744.4900756)}; 
        \addplot[color of colormap={500},mark=pentagon*, opacity=0.5] coordinates {(3.4718802476320083, 529.3156834)}; 
        \addplot[color of colormap={600},mark=halfdiamond*, opacity=0.5] coordinates {(15.829095542330027, 737.4040415000001)}; 
        \addplot[color of colormap={700},mark=halfcircle*] coordinates {(13.5720794915648, 835.2898837)}; 
      \end{loglogaxis}
    \end{tikzpicture}
  }
  \subfloat{
    \begin{tikzpicture}
    	\label{fig:n1e1bestmdofs}
      \begin{loglogaxis}
        [ roofline
        , title = {optimization path}
        , xmax=40
        ]
        \drawroofline[avx=false, fma=false]
        \addplot+[only marks, color of colormap={300},mark=*] coordinates {(8.446623473794608, 233.06776259999998)}; \label{pgf:N1_S} 
        \addplot+[only marks, color of colormap={400},mark=square*] coordinates {(8.231038248118242, 724.7770247)}; \label{pgf:N1_SV} 
        \addplot+[only marks, color of colormap={500},mark=pentagon*] coordinates {(3.4718802476320083, 529.3156834)}; \label{pgf:N1_SVI} 
        \addplot+[only marks, color of colormap={600},mark=halfdiamond*] coordinates {(15.829095542330027, 737.4040415000001)}; \label{pgf:N1_SVIC} 
        \addplot+[only marks, color of colormap={700},mark=halfcircle*] coordinates {(13.5720794915648, 835.2898837)}; \label{pgf:N1_SVICT} 
        \draw[a]                             (axis cs:8.446623473794608, 233.06776259999998) -- node[n,pos=0.3]{V} (axis cs:8.231038248118242, 724.7770247);
        \draw[a]                             (axis cs:8.231038248118242, 724.7770247)        -- node[n        ]{I} (axis cs:3.4718802476320083, 529.3156834);
        \draw[a, bend right]                 (axis cs:3.4718802476320083, 529.3156834)       to node[n,pos=0.3]{C} (axis cs:15.829095542330027, 737.4040415000001);
        \draw[a, out=30, in=60, looseness=7] (axis cs:15.829095542330027, 737.4040415000001) to node[n,pos=0.3]{T} (axis cs:13.5720794915648, 835.2898837);
      \end{loglogaxis}
    \end{tikzpicture}
  }
\vspace{-0.2cm}
  \subfloat{
    \begin{tikzpicture}
    	
      \begin{axis}
        [ performance
        , symbolic y coords = {N1_SVIT, N1_SVIfQT, N1_SVIfQ, N1_SVICT, N1_SVIC, N1_SVI, N1_SVICfQT, N1_SVICfQ,  N1_SVIfQ, N1_SVIT, N1_SVIfQT}
        , ytick = { N1_SVICT, N1_SVIC, N1_SVI, N1_SVICfQT, N1_SVICfQ, N1_SVIfQ, N1_SVIfQT,  N1_SVIT}
        , yticklabels = { N1\_SVICT, N1\_SVIC, N1\_SVI, N1\_SVICfQT, N1\_SVICfQ, N1\_SVIT, N1\_SVIfQT, N1\_SVIfQ}
        ,fill=.!.10
        ]
         \addplot[color of colormap={100},fill=.!30,mark=halfsquare*] coordinates {(1053.8671480144405,N1_SVIT)}; 
           \addplot[color of colormap={0},fill=.!30,mark=halfsquare right*] coordinates {(1080.716049382716,N1_SVIfQT)}; 
        \addplot[color of colormap={200},fill=.!30,mark=halfsquare left*] coordinates {(666.2135920852361,N1_SVIfQ)}; 
         \addplot[color of colormap={700},fill=.!30,mark=halfcircle*] coordinates {(2109.4785542168675,N1_SVICT)}; 
         \addplot[color of colormap={600},fill=.!30,mark=halfdiamond*] coordinates {(1574.0194244604318,N1_SVIC)}; 
        \addplot[color of colormap={500},fill=.!30,mark=pentagon*] coordinates {(1095.1291614518148,N1_SVI)}; 
           \addplot[color of colormap={400},fill=.!30,mark=diamond*] coordinates {(1425.4778501628664,N1_SVICfQT)}; 
        
        \addplot[color of colormap={300},fill=.!30,mark=triangle*] coordinates {(733.3352596314908,N1_SVICfQ)}; 
      
             \end{axis}
    \end{tikzpicture}
  }
  \subfloat{
    \sisetup{round-mode=figures}
    \begin{tikzpicture}
    	\label{fig:n1e1histmdofs}
      \begin{axis}
        [ performance
        , symbolic y coords = {N1_SVICT, N1_SVIC, N1_SVI, N1_SV, N1_S}
        , ytick = {N1_S, N1_SV, N1_SVI, N1_SVIC, N1_SVICT}
        , yticklabels = {N1\_S, N1\_SV, N1\_SVI, N1\_SVIC, N1\_SVICT} 
        ]
        \addplot[color of colormap={300},fill=.!30,mark=*] coordinates {(191.94211795658845,N1_S)}; 
        \draw[a, bend left] (axis cs:191.94211795658845,N1_S) to node[n, anchor=south west]{$\num{2.9554326514238656}\times$} (axis cs:567.2720025923526,N1_SV);
        \addplot[color of colormap={400},fill=.!30,mark=square*] coordinates {(567.2720025923526,N1_SV)}; 
        \draw[a, bend left] (axis cs:567.2720025923526,N1_SV) to node[n, anchor=south west]{$\num{1.9305186161968682}\times$} (axis cs:1095.1291614518148,N1_SVI);
        \addplot[color of colormap={500},fill=.!30,mark=pentagon*] coordinates {(1095.1291614518148,N1_SVI)}; 
        \draw[a, bend left] (axis cs:1095.1291614518148,N1_SVI) to node[n, anchor=south west]{$\num{1.43729112497904}\times$} (axis cs:1574.0194244604318,N1_SVIC);
        \addplot[color of colormap={600},fill=.!30,mark=halfdiamond*] coordinates {(1574.0194244604318,N1_SVIC)}; 
        \draw[a, bend left] (axis cs:1574.0194244604318,N1_SVIC) to node[n, anchor=south west]{$\num{1.3401858461435374}\times$} (axis cs:2109.4785542168675,N1_SVICT);
        \addplot[color of colormap={700},fill=.!30,mark=halfcircle*] coordinates {(2109.4785542168675,N1_SVICT)}; 
      \end{axis}
    \end{tikzpicture}
  }
\vspace{-0.2cm}
  \caption{\textbf{Left column}: Roofline (top) and MDoF/s (bottom) of the best
  optimization combinations for the weak form \cref{eq:n1e1curlcurlplusmass}.
  \textbf{Right column}: Roofline (top) and MDoF/s (bottom) with speed-ups of the
  intermediate operators during the optimization process to obtain \inst{N1_SVICT}. The accumulated speed-up is \num{11}$\times$.}
  \label{fig:n1roof}
  \vspace{-0.75cm}
\end{figure}

\cref{eq:n1e1curlcurlplusmass} operators exhibit vastly superior performance  than
those from \cref{eq:p1diff} with up to \qty{62}{\%} of the AVX FMA peak performance and
2.1 GDoF/s (\cref{fig:n1roof}, left column).
The integrand is more compute-intensive, placing the best operators generally
at higher arithmetic intensities.

The standard optimizations show a familiar pattern in terms of performance, mirroring what we observed for \cref{eq:p1diff}, as visible in \cref{fig:n1roof}, right column: a sharp performance spike from vectorization (V), followed by a shift towards memory boundedness by drawing loop invariants (I) and then a return to the compute-bound region after applying the cubes loop strategy (C). These optimizations culminate in operator \inst{N1_SVIC}.
At this point, tabulation (T) substantially reduces the computational workload and enhances performance by nearly 500 MDoF/s (\cref{fig:n1roof}, lower right).
Two tables are assembled, one for the curl-curl and mass part, respectively.
This is necessary due to the two spatially-varying coefficients in the weak form.
Combining unrolled quadrature loops and tabulation emerges as the most effective
set of optimizations with an accumulated 11$\times$ speed-up in \inst{N1_SVICT}.

\subsection{Variable-Coefficient Diffusion}

A variable coefficient diffusion operator $\nabla \cdot (k(\genVec{x}) \nabla )$ using $\trialspacena = \testspacena = \coeffspacena = \quadPoly(\microGrid{\Omega})$ represents more compute-intense operators.
This discretization leads to $8.9 \times 10^9$ \ac{DoF}s on the test cube. A similar form arises from the
Taylor-Hood $(\quadPoly-\linPoly)$ discretization of the Stokes equation in the case of
a spatially-varying viscosity~\cite{Bauer:2020:TerraNeoMantleConvection}. The
local matrix is
\begin{align}
	\label{eq:p2divkgrad}
	\genLocMicroOp =\locMat{\intMicroElem k(\mathbf{x}) (\nabla \trialbasis^i \cdot \nabla \testbasis^j)}{\iset{\quadPoly}_{T_m}}{\iset{\quadPoly}_{T_m}}. \tag{\textbf{P2V}}
\end{align}

\begin{figure}
  \subfloat{
    \begin{tikzpicture}
      \begin{loglogaxis}
        [ roofline
        , title = {optimization search}
        , xmax=70
        ]
        \drawroofline[xmax=70, angle=39, avx=false, fma=false]

        \addplot[color of colormap={0},mark=halfcircle*] coordinates {(10.749119607424053, 604.3305144)}; 
        \addplot[color of colormap={100},mark=halfsquare right*] coordinates {(48.73394224977833, 500.8231735)}; 
        \addplot[color of colormap={200},mark=halfsquare left*] coordinates {(3.198898476950352, 467.1114737)}; 
        \addplot[color of colormap={300},mark=triangle*] coordinates {(14.955834911241029, 522.3237552)}; 
        \addplot+[color of colormap={400},mark=diamond*] coordinates {(2.8694761136230373, 437.8518032)}; 
        \addplot[color of colormap={500},mark=pentagon*] coordinates {(13.26081294333611, 621.4512714)}; 
        \addplot[color of colormap={600},mark=halfdiamond*] coordinates {(14.688294374009448, 676.4475324)}; 
        \addplot[color of colormap={700},mark=halfcircle*, opacity=0.5] coordinates {(3.1232578678836336, 465.66670270000003)}; 
      \end{loglogaxis}
    \end{tikzpicture}
  }
  \subfloat{
    \begin{tikzpicture}
      \begin{loglogaxis}
        [ roofline
        , title = {optimization path}
        , xmax=70
        ]
        \drawroofline[xmax=70, angle=39, avx=false, fma=false, peak=false]

        \addplot+[only marks, color of colormap={100},mark=*] coordinates {(42.33963040831896, 172.77707420000002)}; \label{pgf:P2V} 
        \addplot+[only marks, color of colormap={200},mark=square*] coordinates {(26.46401099530422, 180.68293979999999)}; \label{pgf:P2V_S} 
        \addplot+[only marks, color of colormap={300},mark=star] coordinates {(26.03240467840423, 596.8998418000001)}; \label{pgf:P2V_SV} 
        \addplot+[only marks, color of colormap={700},mark=oplus] coordinates {(9.011814392338005, 747.6681099000001)}; \label{pgf:P2V_SVU} 
        \addplot+[only marks, color of colormap={400},mark=diamond*] coordinates {(2.8694761136230373, 437.8518032)}; \label{pgf:P2V_SVUI} 
        \addplot+[only marks, color of colormap={500},mark=pentagon*] coordinates {(13.26081294333611, 621.4512714)}; \label{pgf:P2V_SVUIC} 
        \addplot+[only marks, color of colormap={600},mark=halfdiamond*] coordinates {(14.688294374009448, 676.4475324)}; \label{pgf:P2V_SVUICT} 

        \draw[a] (axis cs:42.33963040831896, 172.7770742)  -- node[n,yshift=-5]{S} (axis cs:26.46401099530422, 180.6829398);
        \draw[a] (axis cs:26.46401099530422, 180.6829398)  -- node[n]{V}           (axis cs:26.03240467840423, 596.8998418);
        \draw[a, bend right] (axis cs:26.03240467840423, 596.8998418)  to node[n,pos=0.4]{U}   (axis cs:9.011814392338005, 747.6681099);
        \draw[a] (axis cs:9.011814392338005, 747.6681099)  -- node[n]{I}           (axis cs:2.8694761136230373, 437.8518032);
        \draw[a] (axis cs:2.8694761136230373, 437.8518032) -- node[n,pos=0.6]{C}           (axis cs:13.26081294333611, 621.4512714);
        \draw[a, out=260, in=290, looseness=9] (axis cs:13.26081294333611, 621.4512714) to node[n,anchor=north east]{T} (axis cs:14.688294374009448, 676.4475324);
      \end{loglogaxis}
    \end{tikzpicture}
  }
\vspace{-0.2cm}
  \subfloat{
    \begin{tikzpicture}
      \begin{axis}
        [ performance
        , symbolic y coords = {P2V_SVUIT, P2V_SVUICT, P2V_SVUIC, P2V_SVUI, P2V_SVUICfQT, P2V_SVUIfQT, P2V_SVUICfQ, P2V_SVUIfQ}
        , ytick = {P2V_SVUIfQ, P2V_SVUICfQ, P2V_SVUIfQT, P2V_SVUICfQT, P2V_SVUI, P2V_SVUIC, P2V_SVUICT, P2V_SVUIT}
        , yticklabels = {P2V\_SVUIfQ, P2V\_SVUICfQ, P2V\_SVUIfQT, P2V\_SVUICfQT, P2V\_SVUI, P2V\_SVUIC, P2V\_SVUICT, P2V\_SVUIT}
        ]
        \addplot[color of colormap={0},fill=.!30,mark=halfcircle*] coordinates {(303.33911987860387,P2V_SVUIfQ)}; 
        \addplot[color of colormap={100},fill=.!30,mark=halfsquare right*] coordinates {(250.41684210526313,P2V_SVUICfQ)}; 
        \addplot[color of colormap={200},fill=.!30,mark=halfsquare left*] coordinates {(707.5650849858357,P2V_SVUIfQT)}; 
        \addplot[color of colormap={300},fill=.!30,mark=triangle*] coordinates {(889.6163701067617,P2V_SVUICfQT)}; 
        \addplot[color of colormap={400},fill=.!30,mark=diamond*] coordinates {(866.7648742411102,P2V_SVUI)}; 
        \addplot[color of colormap={500},fill=.!30,mark=pentagon*] coordinates {(1212.4271844660193,P2V_SVUIC)}; 
        \addplot[color of colormap={600},fill=.!30,mark=halfdiamond*] coordinates {(1294.7398963730568,P2V_SVUICT)}; 
        \addplot[color of colormap={700},fill=.!30,mark=halfsquare*] coordinates {(848.2376910016977,P2V_SVUIT)}; 
      \end{axis}
    \end{tikzpicture}
  }
  \subfloat{
    \sisetup{round-mode=figures}
    \begin{tikzpicture}
      \begin{axis}
        [ performance
        , xmax = 1500
        , symbolic y coords = {P2V_SVUICT, P2V_SVUIC, P2V_SVUI, P2V_SVU, P2V_SV, P2V_S, P2V}
        , ytick = {P2V, P2V_S, P2V_SV, P2V_SVU, P2V_SVUI, P2V_SVUIC, P2V_SVUICT}
        , yticklabels = {P2V, P2V\_S, P2V\_SV, P2V\_SVU, P2V\_SVUI, P2V\_SVUIC, P2V\_SVUICT}
        ]
        \addplot[color of colormap={100},mark=*] coordinates {(22.310554513795875,P2V)}; 
        
        \draw[a, bend left] (axis cs:22.310554513795875,P2V) to node[n, anchor=south west]{$\num{1.6709024191942565}\times$} (axis cs:37.278759510666866,P2V_S);
        \addplot[color of colormap={200},fill=.!30,mark=square*] coordinates {(37.278759510666866,P2V_S)}; 
        \draw[a, bend left] (axis cs:37.278759510666866,P2V_S) to node[n, anchor=south west]{$\num{3.23051589555342}\times$} (axis cs:120.42962516572254,P2V_SV);
        \addplot[color of colormap={300},fill=.!30,mark=star] coordinates {(120.42962516572254,P2V_SV)}; 
        \draw[a, bend left] (axis cs:120.42962516572254,P2V_SV) to node[n, anchor=south west]{$\num{3.8498345955430464}\times$} (axis cs:463.6341372912801,P2V_SVU);
        \addplot[color of colormap={700},fill=.!30,mark=oplus] coordinates {(463.6341372912801,P2V_SVU)}; 
        \draw[a, bend left] (axis cs:463.6341372912801,P2V_SVU) to node[n, anchor=south west]{$\num{1.869501843209965}\times$} (axis cs:866.7648742411102,P2V_SVUI);
        \addplot[color of colormap={400},fill=.!30,mark=diamond*] coordinates {(866.7648742411102,P2V_SVUI)}; 
        \draw[a, bend left] (axis cs:866.7648742411102,P2V_SVUI) to node[n, anchor=south west]{$\num{1.3987959370498848}\times$} (axis cs:1212.4271844660193,P2V_SVUIC);
        \addplot[color of colormap={500},fill=.!30,mark=pentagon*] coordinates {(1212.4271844660193,P2V_SVUIC)}; 
        \draw[a, bend left] (axis cs:1212.4271844660193,P2V_SVUIC) to node[n, anchor=south west]{$\num{1.067890849827233}\times$} (axis cs:1294.7398963730568,P2V_SVUICT);
        \addplot[color of colormap={600},fill=.!30,mark=halfdiamond*] coordinates {(1294.7398963730568,P2V_SVUICT)}; 
      \end{axis}
    \end{tikzpicture}
  }
\vspace{-0.3cm}
  \caption{\textbf{Left column}: Roofline (top) and MDoF/s (bottom) of the best
  optimization combinations for the weak form \cref{eq:p2divkgrad}.
  \textbf{Right column}: Roofline (top) and MDoF/s (bottom) with speed-ups of the
  intermediate operators 
   during the optimization process to obtain \inst{P2V_SVUICT}. The accumulated speed-up is \num{58}$\times$.}
  \label{fig:p2vroof}
  \vspace{-0.75cm}
\end{figure}

The left column of \cref{fig:p2vroof} shows that a similar set of optimizations as for \cref{eq:n1e1curlcurlplusmass}, including the standard optimizations, unrolled quadrature loops and tabulation proves most effective also for \cref{eq:p2divkgrad}. Operator \inst{P2V_SVUICT} reaches 1.3 GDoF/s and \qty{50}{\%} AVX FMA peak performance.

We observe the optimization path leading up to \inst{P2V_SVUICT} in \cref{fig:p2vroof}, right column, where additional optimizations are applied together with vectorization (V), moving loop invariants (I) and the cubes loop strategy (C). The latter cause the familiar upwards, left and right shifts in the roofline model.

The integrand of \cref{eq:p2divkgrad} is a polynomial of fourth degree due to the quadratic test,
trial and coefficient space.
Evaluating the integrand on 11 quadrature points defined by Xiao-Gimbutas rule~\cite{Xiao:2010:Quad} bloats the \ac{AST}, preventing the \ac{CSE} to detect symmetry.
Explicitly exploiting symmetry in the
\fogshort (S) eliminates 45 of the 100 entries in
the local matrix (for 10 shape functions in test and trial space, respectively).

\cref{eq:p2divkgrad} offers the option to under-integrate (U) due to the difference
in polynomial degree in the integrand and the shape functions.
By applying the Xiao-Gimbutas rule of second order instead of fourth order,
the number of quadrature points reduces from 11 to 4, with a corresponding
speed-up.
We verified that the proper convergence rate is still achieved
when solving analytical test cases on multiple refinement levels using
\inst{P2V_SVU}.
Note that using a quadrature rule merely exact for linear polynomials destroys
the convergence order.

By various optimizations reducing the arithmetic intensity, the \fogshort has
transformed the compute-intense \inst{P2V}, placed deep in the compute-bound
region, into the memory-bound \inst{P2V_SVUI}.
The memory volume becomes the operator's bottleneck, so reducing it by the cubes loop strategy
provides another significant speed-up, accumulating to $58\times$ in the
final operator \inst{P2V_SVUICT}.

\subsubsection{Instruction Boundedness}
\label{sec:InstBoundedness}

So far, unrolling the quadrature loops and applying \ac{CSE} to the resulting
statements has proved effective, entering the fastest operators for all weak forms.
However, during experiments we observed that as compute-intensity of the operators increases, unrolling can cause a non-obvious bottleneck.

We generate \inst{P2V_VI}, which unrolls the quadrature loop, does not tabulate and uses a
quadrature rule exact for polynomials of fourth degree.
To reveal the bottleneck, we require an L2-roofline where the memory roof and arithmetic
intensity are computed using the L2-cache bandwidth.
We do not under-integrate here for demonstration purpose, as it would obscure
the analysis of the bottleneck.

\begin{figure}
	\subfloat{
		\begin{tikzpicture}			
			\label{fig:p2v-insts-roof}
			\begin{loglogaxis}
				[ roofline
				, title = {L2 roofline}
				, xmax=70
				,  cycle multiindex* list = {
					[samples of colormap = 5]\nextlist
					mark list*\nextlist
				}
				]
				\drawroofline[ltwo, xmax=70, angle=51, avx=false, fma=false]
				
				\addplot+[only marks
				] coordinates {(0.3626820824089631, 586.0815513000001)};
				\label{pgf:P2V_VI} 
				\addplot+[only marks
				] coordinates {(23.305169817045094, 488.4106551)};    \label{pgf:P2V_VIfQ} 
				\draw[a] (axis cs:0.3626820824089631, 586.0815513000001) to node[n]{fQ} (axis cs:23.305169817045094, 488.4106551);
				
			\end{loglogaxis}
		\end{tikzpicture}
	}
	\subfloat{
			\begin{tikzpicture}			
			\label{fig:p2v-precomp-roof}
			\begin{loglogaxis}
				[ roofline
				, title = {comparison with precomputing}
				, xmax=20
					,  cycle multiindex* list = {
					[samples of colormap = 5]\nextlist
					mark list*\nextlist
				}
				]
				\drawroofline[xmin=0.2,xmax=20, angle=31,avxfma=false, avx=false, fma=false]
				\pgfplotsset{cycle list shift = -1}
				\addplot+[only marks] coordinates {(0.3465563780040332, 39)};
				\label{pgf:P2V_P} 
				\addplot+[only marks] coordinates {(7.1, 520)};    \label{pgf:P2V_VIfQT} 
				
			\end{loglogaxis}
		\end{tikzpicture}
	}
\\
\subfloat{
	\hspace{0.75cm}
	\resizebox{5cm}{!}{%
		\begin{tabular}{cc}
			\toprule
			& L2inst [GB] (\% of L2) \\
			\midrule
			P2V\_VI	\ref{pgf:P2V_VI} &   $1.8 \times 10^3 \ (78)$ \\
			P2V\_VIfQ \ref{pgf:P2V_VIfQ} & $1.3 \ (4)$\\
			\bottomrule
		\end{tabular}
	}
}
\subfloat{
		\resizebox{6.5cm}{!}{%
		\begin{tabular}{cccc}
			\toprule
			& FLOP/$\micro$ & mem [GB] & MDoF/s \\
			\midrule
			P2V\_P	\ref{pgf:P2V_P} & $210$ & $66$ & 166\\
			P2V\_VIfQT \ref{pgf:P2V_VIfQT} &$1.6 \times 10^3$  & $16$ & 421\\
			\bottomrule
		\end{tabular}
	}
}
	\caption{\textbf{Left column}: L2 roofline and L2 cache volume induced by loading instructions. The fraction of the L2 instruction volume with respect to the overall L2 cache volume (load and store) is annotated in parentheses. \textbf{Right column}: Roofline, FLOPS per element, main memory volume and performance compared against precomputation of local element matrices.}
	\label{fig:inst-precom}
	\vspace{-20pt}
\end{figure}
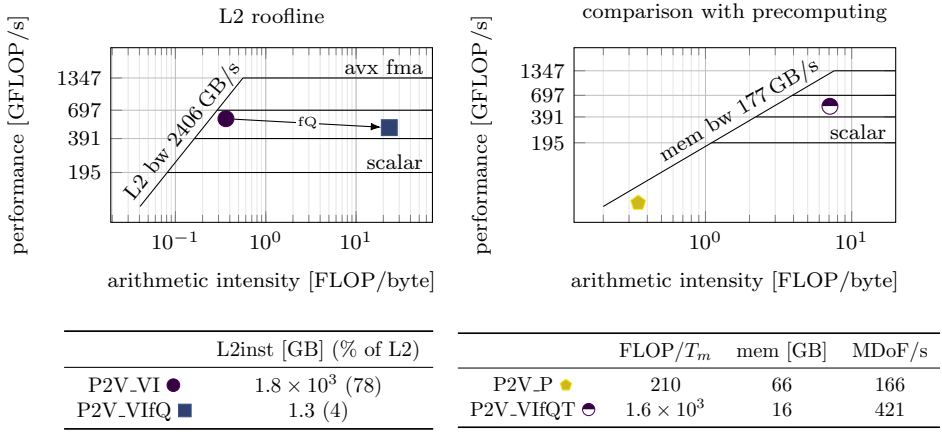

The L2 roofline (\cref{fig:inst-precom}, upper left) shows that the operator \inst{P2V_VI} lies very close to the
L2-bandwidth roof.
It suffers from a very large L2 cache volume, which does not stem from data
transfers but instructions being loaded from L2 to L1 cache
(\cref{fig:inst-precom}, lower left).
\inst{P2V_VI} is \emph{instruction-bound}.
Unrolling the quadrature loops with many points and a more involved weak form
bloats the number of statements in the kernel body.
Consequently, a large amount of distinct instructions has to be fetched,
causing the high L2 cache volume.
Simply not unrolling the quadrature loop reduces the number of statements in the
kernel body and the instruction traffic to negligible amounts, alleviating the
instruction-boundedness with \inst{P2V_VIfQ}.

\subsubsection{Comparison with Precomputation of Local Element Matrices}
Given that sufficient memory is available, the local matrix for each element of the grid can be precomputed, loaded and applied during the loop~\cite{Carey:1986:CANM, Carey:1988:elembyelem}.
We compare our generated operators against this precomputation technique for completeness and because it seems to be commonly applied to variable-coefficient problems in \ac{FE} simulations. We establish a fair comparison against our implementation of precomputation by not exploiting symmetry and using the sawtooth loop strategy in both operators.

The right column of \cref{fig:inst-precom} demonstrates the drawback of precomputing local matrices:
although \inst{P2V_P} does relatively few \acp{FLOP}, it suffers from a large memory load from the local matrices,
which renders it extremely memory-bound. For reference, the \cref{eq:p1diff} operator with the lowest arithmetic intensity, \inst{P1_SVIfQ}, lies at \num{1.5} FLOP/byte, while \inst{P2V_P} places at just \num{0.3} FLOP/byte.
\inst{P2V_VIfQT} on the other hand assembles on the fly, does a magnitude more computation, but places well

into the machine balance. Thereby, it runs $2.5\times$ faster.

\section{Scaling Curl-Curl to a Trillion DoFs}
\label{sec:ExtremeScale}
Finally, we demonstrate the weak scalability of the generated operators within the \hyteg framework.

The (homogeneous) curl-curl problem
\begin{equation}
	\label{eq:hyteg-demo-curl-curl}
	\begin{aligned}
		\alpha\, \genVec{curl}\, \genVec{curl}\, \genVec{u} + \beta \genVec{u} &= \genVec{f} && \text{in}\ \Omega, \\
		\genVec{u} \times \genVec{n} &= 0 && \text{on}\ \partial \Omega,
	\end{aligned}
\end{equation}
in three dimensions, with given $\genVec{f} \in \genVec L^2(\Omega)$, and $\alpha,\ \beta \in L^2(\Omega)$, arises from Maxwell's
equations in electromagnetic wave scattering problems~\cite{Hipmair:2009:MGHcurl}.
We discretize the solution $\genVec{u}$ and \ac{RHS} $\genVec{f}$ with first order, first kind Nédélec elements ($\nedelec$).  
The corresponding matrix-free operator we use is \inst{N1_SVIC} from \cref{sec:CCPM}, implementing the weak form \cref{eq:n1e1curlcurlplusmass}.

To assess the weak scalability, we solve \cref{eq:hyteg-demo-curl-curl} on the unit cube with an equally growing number of processes and coarse grid elements.
In all runs there is a total of eight levels in the grid hierarchy.
For simplicity, the coefficients $\alpha(\genVec x)$ and $\beta(\genVec x)$ equal \num{1} for all $\genVec x$.
Nonetheless, the operators treat them as if they are spatially varying.

The system is solved using matrix-free \ac{FMG} with a single V(1,1) cycle on each level.
The hybrid smoother published in~\cite{Hipmair:2009:MGHcurl} is used due to the non-elliptic nature of the bilinear form.
In short, smoothing is split into two sub-steps.
While a standard smoother in $\nedelec$ reduces error components in $\mathcal N(\genVec{curl})^\perp$,
the nullspace is handled by relaxing on Poisson's equation in potential space, which is discretized using $\linPoly$ \acp{FE}.
We choose Chebyshev smoothers of order~\num{2}~\cite{Adams:2003:ParallelMultigridSmoothing,Baker:2011:MultigridSmoothersUltraparallel} in both spaces.
This means that one hybrid smoothing step requires in total three matrix-vector products in $\nedelec$, two matrix-vector products in $\linPoly$, and two transfer operations between the spaces.
Note that two additional $\linPoly$ vectors must be allocated.
The $\linPoly$ operator is generated with the same set of optimizations as the $\nedelec$ operator (SVIC).
On the coarsest grid the matrix is assembled, and PETSc's~\cite{PETSc:1997:Efficient} SOR preconditioned CG solver solves the system up to a relative residual reduction of $\num{e-3}$.

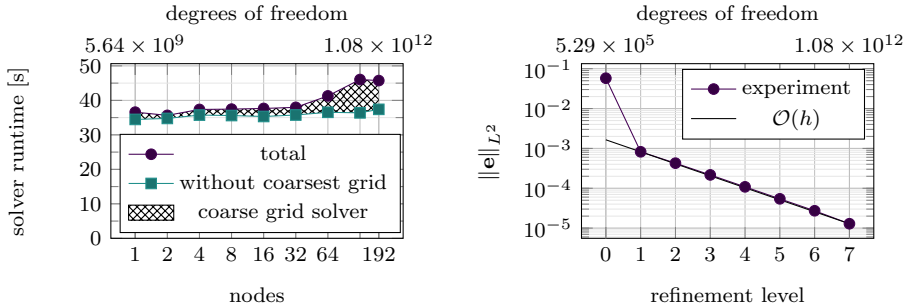
\begin{figure}
	\centering
	\subfloat{
		\begin{tikzpicture}
  \sisetup{exponent-mode=scientific, round-mode=places}
  \begin{axis}
    [ width  = 0.42\textwidth
    , height = 0.3\textwidth
    , xmode = log
    , ymin  = 0
    , xtick = {1,2,4,8,16,32,64,128,192}
    , xticklabels = {$1$,$2$,$4$,$8$,$16$,$32$,$64$,,$192$}
    , extra x ticks = {1,192}
    , extra x tick labels = {$\num{5643437824}$,$\num{1082539386880}$}
    , extra x tick style = {xticklabel pos = upper}
    , ytick distance = 10
    , minor y tick num = 1
    , xlabel = {nodes}
    , ylabel = {solver runtime [\unit{s}]}
    , title  = {degrees of freedom}
    , title style = {at={(0.5,1.08)}}
    , legend pos = south west
    , grid = both
    , mark options = {scale = 1}
     ,  cycle multiindex* list = {
    	[samples of colormap = 3]\nextlist
    	mark list*\nextlist
    }
    ]
    \addplot+[name path=all] table[x=nodes, y=timeMaxAllLvl] {data/weak-scaling.dat};
    \addlegendentry{total};

    \addplot+[name path=cgs] table[x=nodes, y expr=\thisrow{timeMaxAllLvl}-\thisrow{timeAvgCgs}] {data/weak-scaling.dat};
    \addlegendentry{without coarsest grid};

    \addplot[pattern=crosshatch] fill between[of=cgs and all];
    \addlegendentry{coarse grid solver};
  \end{axis}
\end{tikzpicture}
	}
	\subfloat{
		\centering
		\begin{tikzpicture}
  \sisetup{exponent-mode=scientific, round-mode=places}
  \begin{axis}[ width  = 0.42\textwidth
              , height = 0.3\textwidth
              , xlabel = {refinement level}
              , ylabel = {$\|\genVec e\|_{L^2}$}
              , title  = {degrees of freedom}
              , title style = {at={(0.5,1.08)}}
              , xtick distance   = 1
              , ytick distance   = 10^1
              , extra x ticks = {0,7}
              , extra x tick labels = {$\num{528848}$,$\num{1082539386880}$}
              , extra x tick style = {xticklabel pos = upper}
              , ymode            = log
              , grid             = both
              , legend pos       = north east
              , mark options = {scale = 1}
              ,  cycle multiindex* list = {
              	[samples of colormap = 3]\nextlist
              	mark list*\nextlist
              }
              ]
    \addplot table[y = L2error] {data/curlcurl_sng2_2024-03-01T18-53-28_nodes_0192_cubes_16-12-12_ref_1_lvl_7.table.dat};
    \addlegendentry{experiment}

    \addplot[black, domain = 0:7] { 1.2853203125e-5 * 2^(7-x) };
    \addlegendentry{$\mathcal{O}(h)$}
  \end{axis}
\end{tikzpicture}
	}
\vspace{-0.3cm}
	\caption{%
		\textbf{Left}: Weak scaling of \hyteg's \ac{FMG} solver is almost perfect from \num{1} to \num{32} nodes. In larger runs the coarse grid solver (not optimized in this work) starts having significant impact. The remaining parts of the multigrid solver keep scaling well.
		\textbf{Right}: $L^2$-error grid convergence for the largest case (\num{192} nodes, \num{21504} parallel processes, \num{1.08e12} \acp{DoF} on level \num{7}).}
	\label{fig:weak-scaling-L2}
	\vspace{-1cm}
\end{figure}

The experiment is performed on the SuperMUC-NG Phase 2 cluster~\cite{LRZ:2024:SNG2}.
Each node comprises two Intel Xeon Platinum 8480+ (Sapphire Rapids) CPUs with \num{56} cores each and \qty{512}{GB} DDR5 memory.
Note that this architecture differs from the one used in the previous section.
Weak scaling results from \num{1} to \num{192} nodes (\num{112} to \num{21504} cores) are summarized in \cref{fig:weak-scaling-L2} (left).
Near perfect scaling is seen up to \num{32} nodes.
Starting with \num{64} nodes, the coarse grid solver starts to have a significant impact on the overall solve time.
Presumably this is due to, first, the growing coarse grid size and suboptimal scaling of the CG-solver, and second, the low computational complexity compared to the high communication cost.
It is expected that for larger problems the coarse grid solver will have a significant
impact on the time to solution. In previous work \cite{Buttari:2022:BLR} it was shown how advanced sparse direct solvers can be employed to alleviate this problem. Alternatives include using parallel AMG methods, such as, e.g., hypre \cite{Falgout:2002:hypreAL} as scalable coarse grid solvers. However, the systematic analysis of these alternatives is beyond the scope of the present article and will be studied in future work.
On the other hand, the remaining part of the solver scales very well, even up to a trillion (\num{e12}) \acp{DoF}.

To assert that the solver finds the correct solution, the $L^2$-error against the known manufactured solution is computed.
\Cref{fig:weak-scaling-L2} (right) shows that the error reduces at the expected linear rate~\cite{Zhong:2009:OptimalErrorEstimates}.

\section{Conclusion}
\label{sec:conclusion}

The presented code generator is a powerful tool for the rapid development of matrix-free \ac{FE} operators. A wide range of optimizations can be applied ``at the press of a button'' to any weak form and a wide array of matrix-free operations.
Notable operations include matrix-vector products, the assembly of an inverse diagonal for Chebychev smoothers and diagonally lumped operators.
Code generation is a vital tool to ensure sustainable high performance even for future applications and hardware architectures.

Using the \fogshort, high node-level performance can be obtained quickly even for complex problems on extreme scales, e.g., the simulation of whole-planet Earth mantle convection. In such simulations, the reduction in runtime resulting from \fogshort's optimizations ultimately saves huge amounts of energy and money.

Through rigorous and systematic analysis surprisingly large speed-ups over existing production codes can be achieved.
We found that traditional tools from performance engineering like the roofline model are essential for the identification of bottlenecks.
Additionally, they make it possible to classify codes as ``slow'' or ``fast''.
In combination with code generation, a multitude of optimizations and combinations thereof can be evaluated with low effort to find an optimum.

Future extensions of the \fogshort include support for curvilinear boundaries,
discontinuous Galerkin methods, vector function spaces, surrogate operators and multigrid grid-transfer operators.
The abstract intermediate representation during generation time enables the development of
additional backends for the automated generation of kernels for accelerators,
particularly GPUs.

\section*{Acknowledgments}
The authors gratefully acknowledge funding through the joint BMBF project CoMPS\footnote{\url{https://gauss-allianz.de/en/project/title/CoMPS}} (grant \texttt{16ME0647K}).
The authors would like to thank the NHR-Verein e.V.\footnote{\url{https://www.nhr-verein.de}} for supporting this work/project within the NHR Graduate School of National High Performance Computing (NHR).
The authors gratefully acknowledge the scientific support and HPC resources provided by the Erlangen National High
Performance Computing Center (NHR@FAU) of the Friedrich-Alexander-Universität Erlangen-Nürnberg (FAU).
NHR funding is provided by federal and Bavarian state authorities.
NHR@FAU hardware is partially funded by the German Research Foundation (DFG) – 440719683.
The authors gratefully acknowledge the Gauss Centre for Supercomputing e.V.\footnote{\url{https://www.gauss-centre.eu}} for funding this project by providing friendly user access during the pilot operation of the GCS Supercomputer SuperMUC-NG Phase 2 at Leibniz Supercomputing Centre\footnote{\url{https://www.lrz.de}}.

\bibliographystyle{siamplain}
\bibliography{references,history}

\begin{thebibliography}{10}

\bibitem{Adams:2003:ParallelMultigridSmoothing}
{\sc M.~Adams, M.~Brezina, J.~Hu, and R.~Tuminaro}, {\em Parallel multigrid
  smoothing: Polynomial versus {{Gauss}}\textendash{{Seidel}}}, Journal of
  Computational Physics, 188 (2003), pp.~593--610,
  \url{https://doi.org/10.1016/S0021-9991(03)00194-3}.

\bibitem{Alnaes:2015:FEniCSProjectVersion}
{\sc M.~Aln{\ae}s, J.~Blechta, J.~Hake, A.~Johansson, B.~Kehlet, A.~Logg,
  C.~Richardson, J.~Ring, M.~Rognes, and G.~Wells}, {\em The {{FEniCS}} project
  version 1.5}, Arch. Num. Soft., 3 (2015).

\bibitem{Baker:2011:MultigridSmoothersUltraparallel}
{\sc A.~H. Baker, R.~D. Falgout, T.~V. Kolev, and U.~M. Yang}, {\em Multigrid
  {{Smoothers}} for {{Ultraparallel Computing}}}, SIAM J. Sci. Comput., 33
  (2011), pp.~2864--2887, \url{https://doi.org/10.1137/100798806}.

\bibitem{PETSc:1997:Efficient}
{\sc S.~Balay, W.~D. Gropp, L.~C. McInnes, and B.~F. Smith}, {\em Efficient
  management of parallelism in object oriented numerical software libraries},
  in Modern Software Tools in Scientific Computing, E.~Arge, A.~M. Bruaset, and
  H.~P. Langtangen, eds., Birkh{\"{a}}user Press, 1997, pp.~163--202,
  \url{https://doi.org/10.1007/978-1-4612-1986-6_8}.

\bibitem{Bauer:2019:Pystencils}
{\sc M.~Bauer, J.~Hötzer, D.~Ernst, J.~Hammer, M.~Seiz, H.~Hierl, J.~Hönig,
  H.~Köstler, G.~Wellein, B.~Nestler, and U.~Rüde}, {\em Code generation for
  massively parallel phase-field simulations}, in SC '19: Proceedings of the
  International Conference for High Performance Computing, Networking, Storage
  and Analysis, IEEE Computer Society, 11 2019, pp.~1--32,
  \url{https://doi.org/10.1145/3295500.3356186}.

\bibitem{Bauer:2020:TerraNeoMantleConvection}
{\sc S.~Bauer, H.-P. Bunge, D.~Drzisga, S.~Ghelichkhan, M.~Huber, N.~Kohl,
  M.~Mohr, U.~R{\"u}de, D.~Th{\"o}nnes, and B.~Wohlmuth}, {\em {{TerraNeo}}
  \textemdash{} {{Mantle Convection Beyond}} a {{Trillion Degrees}} of
  {{Freedom}}}, in Softw. {{Exascale Comput}}. - {{SPPEXA}} 2016-2019, H.-J.
  Bungartz, S.~Reiz, B.~Uekermann, P.~Neumann, and W.~Nagel, eds., vol.~136 of
  Lecture Notes in Computational Science and Engineering, {Springer}, 2020,
  pp.~569--610, \url{https://doi.org/10.1007/978-3-030-47956-5_19}.

\bibitem{Bergen:2004:HierarchicalHybridGrids}
{\sc B.~K. Bergen and F.~H{\"u}lsemann}, {\em Hierarchical hybrid grids: Data
  structures and core algorithms for multigrid}, Numer. Linear Algebra Appl.,
  11 (2004), pp.~279--291, \url{https://doi.org/10.1002/nla.382}.

\bibitem{Bey:1995:TetrahedralGridRefinement}
{\sc J.~Bey}, {\em Tetrahedral grid refinement}, Computing, 55 (1995),
  pp.~355--378, \url{https://doi.org/10.1007/BF02238487}.

\bibitem{Brenner:2008:FE}
{\sc S.~C. Brenner and L.~R. Scott}, {\em {The Mathematical Theory of Finite
  Element Methods}}, Springer-Verlag, 2008,
  \url{https://doi.org/10.1007/978-0-387-75934-0}.

\bibitem{Brown:2010:JSC}
{\sc J.~Brown}, {\em Efficient {N}onlinear {S}olvers for {N}odal {H}igh-{O}rder
  {F}inite {E}lements in {3D}}, Journal of Scientific Computing, 45 (2010),
  pp.~48--63, \url{https://doi.org/10.1007/s10915-010-9396-8}.

\bibitem{Buttari:2022:BLR}
{\sc A.~Buttari, M.~Huber, P.~Leleux, T.~Mary, U.~R{\"u}de, and B.~Wohlmuth},
  {\em Block low-rank single precision coarse grid solvers for extreme scale
  multigrid methods}, Numerical Linear Algebra with Applications, 29 (2022),
  \url{https://doi.org/10.1002/nla.2407}.

\bibitem{Carey:1988:elembyelem}
{\sc G.~F. Carey, E.~J. Barragy, R.~T. McLay, and M.~Sharma}, {\em
  Element‐by‐element vector and parallel computations}, Communications in
  Applied Numerical Methods, 4 (1988), pp.~299--307,
  \url{https://api.semanticscholar.org/CorpusID:120879806}.

\bibitem{Carey:1986:CANM}
{\sc G.~F. Carey and B.-N. Jiang}, {\em Element-by-element linear and nonlinear
  solution schemes}, International Journal for Numerical Methods in Biomedical
  Engineering (Formerly: Communications in Numerical Methods in Engineering;
  Communications in Applied Numerical Methods), 2 (1986), pp.~145--153,
  \url{https://doi.org/10.1002/cnm.1630020205}.

\bibitem{Ciarlet:2002:FE}
{\sc P.~G. Ciarlet}, {\em The Finite Element Method for Elliptic Problems},
  Society for Industrial and Applied Mathematics, 2002,
  \url{https://doi.org/10.1137/1.9780898719208}.

\bibitem{Falgout:2002:hypreAL}
{\sc R.~D. Falgout and U.~M. Yang}, {\em hypre: A library of high performance
  preconditioners}, in Computational Science --- ICCS 2002, P.~M.~A. Sloot,
  A.~G. Hoekstra, C.~J.~K. Tan, and J.~J. Dongarra, eds., 2002, pp.~632--641,
  \url{https://doi.org/10.1007/3-540-47789-6_66}.

\bibitem{Gmeiner:2016:QuantitativePerformanceStudy}
{\sc B.~Gmeiner, M.~Huber, L.~John, U.~R{\"u}de, and B.~Wohlmuth}, {\em A
  quantitative performance study for {{Stokes}} solvers at the extreme scale},
  Journal of Computational Science, 17 (2016), pp.~509--521,
  \url{https://doi.org/10.1016/j.jocs.2016.06.006}.

\bibitem{Gmeiner:2015:PerformanceScalabilityHierarchical}
{\sc B.~Gmeiner, U.~R{\"u}de, H.~Stengel, C.~Waluga, and B.~Wohlmuth}, {\em
  Performance and {{Scalability}} of {{Hierarchical Hybrid Multigrid Solvers}}
  for {{Stokes Systems}}}, SIAM J. Sci. Comput., 37 (2015), pp.~C143--C168,
  \url{https://doi.org/10.1137/130941353}.

\bibitem{Hager:2010:IntroductionHighPerformance}
{\sc G.~Hager and G.~Wellein}, {\em Introduction to {{High Performance
  Computing}} for {{Scientists}} and {{Engineers}}}, {CRC Press}, 1~ed., July
  2010, \url{https://doi.org/10.1201/EBK1439811924}.

\bibitem{Ham:2023:Firedrake}
{\sc D.~A. Ham, P.~H.~J. Kelly, L.~Mitchell, C.~J. Cotter, R.~C. Kirby,
  K.~Sagiyama, N.~Bouziani, S.~Vorderwuelbecke, T.~J. Gregory, J.~Betteridge,
  D.~R. Shapero, R.~W. Nixon-Hill, C.~J. Ward, P.~E. Farrell, P.~D. Brubeck,
  I.~Marsden, T.~H. Gibson, M.~Homolya, T.~Sun, A.~T.~T. McRae, F.~Luporini,
  A.~Gregory, M.~Lange, S.~W. Funke, F.~Rathgeber, G.-T. Bercea, and G.~R.
  Markall}, {\em Firedrake User Manual}, Imperial College London and University
  of Oxford and Baylor University and University of Washington, first
  edition~ed., 5 2023, \url{https://doi.org/10.25561/104839}.

\bibitem{Hammer:2017:kernkraft}
{\sc J.~Hammer, J.~Eitzinger, G.~Hager, and G.~Wellein}, {\em Kerncraft: A Tool
  for Analytic Performance Modeling of Loop Kernels}, Springer International
  Publishing, 2017, pp.~1--22,
  \url{https://doi.org/10.1007/978-3-319-56702-0_1}.

\bibitem{Hipmair:2009:MGHcurl}
{\sc R.~Hiptmair}, {\em Multigrid {M}ethod for {M}axwell's {E}quations}, SIAM
  J. Numer. Anal., 36 (1998), pp.~204--225,
  \url{https://doi.org/10.1137/S0036142997326203}.

\bibitem{Hosangadi:2006:polyCSE}
{\sc A.~Hosangadi, F.~Fallah, and R.~Kastner}, {\em Optimizing polynomial
  expressions by algebraic factorization and common subexpression elimination},
  IEEE Transactions on Computer-Aided Design of Integrated Circuits and
  Systems, 25 (2006), pp.~2012--2021,
  \url{https://doi.org/10.1109/TCAD.2006.875712}.

\bibitem{Hughes:1986:stabilizedStokes}
{\sc T.~J. Hughes, L.~P. Franca, and M.~Balestra}, {\em A new finite element
  formulation for computational fluid dynamics: V. circumventing the
  {Babuška-Brezzi} condition: a stable {P}etrov-{G}alerkin formulation of the
  stokes problem accommodating equal-order interpolations}, Computer Methods in
  Applied Mechanics and Engineering, 59 (1986), pp.~85--99,
  \url{https://doi.org/10.1016/0045-7825(86)90025-3}.

\bibitem{Kirby:2006:FFC}
{\sc R.~C. Kirby and A.~Logg}, {\em A compiler for variational forms}, ACM
  Transactions on Mathematical Software, 32 (2006),
  \url{https://doi.org/10.1145/1163641.1163644}.

\bibitem{Kirby:2018:SolverComposition}
{\sc R.~C. Kirby and L.~Mitchell}, {\em Solver composition across the
  pde/linear algebra barrier}, SIAM Journal on Scientific Computing, 40 (2018),
  p.~C76–C98, \url{https://doi.org/10.1137/17m1133208}.

\bibitem{Kohl:2023:FundamentalDataStructures}
{\sc N.~Kohl, D.~Bauer, F.~Böhm, and U.~Rüde}, {\em Fundamental data
  structures for matrix-free finite elements on hybrid tetrahedral grids}, Int.
  J. Parallel Emergent Distrib. Syst., 39 (2024), pp.~51--74,
  \url{https://doi.org/10.1080/17445760.2023.2266875}.

\bibitem{Kohl:2022:SISC}
{\sc N.~Kohl, M.~Mohr, S.~Eibl, and U.~R{\"u}de}, {\em A {M}assively {P}arallel
  {E}ulerian-{L}agrangian {M}ethod for {A}dvection-{D}ominated {T}ransport in
  {V}iscous {F}luids}, SIAM J.~Sci.~Comp., 44 (2022), pp.~C260--C285,
  \url{https://doi.org/10.1137/21M1402510}.

\bibitem{Kohl:2022:TextbookEfficiencyMassively}
{\sc N.~Kohl and U.~R{\"u}de}, {\em Textbook {{Efficiency}}: {{Massively
  Parallel Matrix-Free Multigrid}} for the {{Stokes System}}}, SIAM J. Sci.
  Comput., 44 (2022), pp.~C124--C155, \url{https://doi.org/10.1137/20M1376005}.

\bibitem{Kohl:2019:HyTeGFiniteelementSoftware}
{\sc N.~Kohl, D.~Th{\"o}nnes, D.~Drzisga, D.~Bartuschat, and U.~R{\"u}de}, {\em
  The {{{\emph{HyTeG}}}} finite-element software framework for scalable
  multigrid solvers}, International Journal of Parallel, Emergent and
  Distributed Systems, 34 (2019), pp.~477--496,
  \url{https://doi.org/10.1080/17445760.2018.1506453}.

\bibitem{Kronbichler:2012:CellBasedOA}
{\sc M.~Kronbichler and K.~Kormann}, {\em A generic interface for parallel
  cell-based finite element operator application}, Computers and Fluids, 63
  (2012), pp.~135--147, \url{https://doi.org/10.1016/j.compfluid.2012.04.012}.

\bibitem{Kronbichler:2019:FastMatrixFreeEvaluation}
{\sc M.~Kronbichler and K.~Kormann}, {\em Fast {{Matrix-Free Evaluation}} of
  {{Discontinuous Galerkin Finite Element Operators}}}, ACM Trans. Math.
  Softw., 45 (2019), pp.~1--40, \url{https://doi.org/10.1145/3325864}.

\bibitem{Kronbichler:2018:PerformanceComparisonContinuous}
{\sc M.~Kronbichler and W.~A. Wall}, {\em A {{Performance Comparison}} of
  {{Continuous}} and {{Discontinuous Galerkin Methods}} with {{Fast Multigrid
  Solvers}}}, SIAM J. Sci. Comput., 40 (2018), pp.~A3423--A3448,
  \url{https://doi.org/10.1137/16M110455X}.

\bibitem{LRZ:2024:SNG2}
{\sc {Leibniz-Rechenzentrum (LRZ)}}, {\em Pilot operation {SuperMUC-NG} {P}hase
  2},
  \url{https://doku.lrz.de/pilot-operation-supermuc-ng-phase-2-403079197.html}
  (accessed 2024-03-02).

\bibitem{RuedeKoestler:2020:ExaStencils}
{\sc C.~Lengauer, S.~Apel, M.~Bolten, S.~Chiba, U.~R{\"u}de, J.~Teich,
  A.~Gr{\"o}{\ss}linger, F.~Hannig, H.~K{\"o}stler, L.~Claus, et~al.}, {\em
  Exastencils: Advanced multigrid solver generation}, in Software for Exascale
  Computing - SPPEXA 2016-2019, Cham, 2020, Springer International Publishing,
  pp.~405--452, \url{https://doi.org/10.1007/978-3-030-47956-5_14}.

\bibitem{May:2014:HPCLithospheric}
{\sc D.~May, J.~Brown, and L.~le~pourhiet}, {\em {PTatin3D: High-Performance
  Methods for Long-Term Lithospheric Dynamics}}, in SC '14: Proceedings of the
  International Conference for High Performance Computing, Networking, Storage
  and Analysis, 11 2014, \url{https://doi.org/10.1109/SC.2014.28}.

\bibitem{May:2015:ScalableMatrixfreeMultigrid}
{\sc D.~May, J.~Brown, and L.~Le~Pourhiet}, {\em A scalable, matrix-free
  multigrid preconditioner for finite element discretizations of heterogeneous
  {{Stokes}} flow}, Computer Methods in Applied Mechanics and Engineering, 290
  (2015), pp.~496--523, \url{https://doi.org/10.1016/j.cma.2015.03.014}.

\bibitem{Meurer:2017:Sympy}
{\sc A.~Meurer, C.~P. Smith, M.~Paprocki, O.~\v{C}ert\'{i}k, S.~B. Kirpichev,
  M.~Rocklin, A.~Kumar, S.~Ivanov, J.~K. Moore, S.~Singh, T.~Rathnayake,
  S.~Vig, B.~E. Granger, R.~P. Muller, F.~Bonazzi, H.~Gupta, S.~Vats,
  F.~Johansson, F.~Pedregosa, M.~J. Curry, A.~R. Terrel, v.~Rou\v{c}ka,
  A.~Saboo, I.~Fernando, S.~Kulal, R.~Cimrman, and A.~Scopatz}, {\em Sympy:
  symbolic computing in python}, PeerJ Computer Science, 3 (2017), p.~e103,
  \url{https://doi.org/10.7717/peerj-cs.103}.

\bibitem{Nedelec:1980:NedelecElems}
{\sc J.~C. Nedelec}, {\em Mixed finite elements in $\mathbb{R}^3$}, Numer.
  Math., 35 (1980), pp.~315--341, \url{https://doi.org/10.1007/bf01396415}.

\bibitem{NHR:2022:Fritzonline}
{\sc NHR@FAU}, {\em Fritz {P}arallel {C}luster}.
\newblock
  \url{https://hpc.fau.de/systems-services/documentation-instructions/clusters/fritz-cluster}.
\newblock Accessed: 2024/01/12.

\bibitem{Rudi:2015:ExtremeScaleImplicit}
{\sc J.~Rudi, A.~C.~I. Malossi, T.~Isaac, G.~Stadler, M.~Gurnis, P.~W.~J.
  Staar, Y.~Ineichen, C.~Bekas, A.~Curioni, and O.~Ghattas}, {\em An
  extreme-scale implicit solver for complex pdes: highly heterogeneous flow in
  earth's mantle}, in SC '15: Proceedings of the International Conference for
  High Performance Computing, Networking, Storage and Analysis, 2015,
  pp.~1--12, \url{https://doi.org/10.1145/2807591.2807675}.

\bibitem{Schloemer:2021:Quadpy}
{\sc N.~Schlömer, N.~Papior, D.~Arnold, J.~Blechta, and R.~Zetter}, {\em
  nschloe/quadpy: None (v0.16.10)}.
\newblock Zenodo, \url{https://doi.org/10.5281/zenodo.5541216}.

\bibitem{Stengel:2014:StencilPerfBottlenecks}
{\sc H.~Stengel, J.~Treibig, G.~Hager, and G.~Wellein}, {\em Quantifying
  performance bottlenecks of stencil computations using the
  execution-cache-memory model}, in ICS '15: Proceedings of the 29th ACM on
  International Conference on Supercomputing, 2015, p.~207–216,
  \url{https://doi.org/10.1145/2751205.2751240}.

\bibitem{Sun:2019:Vect}
{\sc T.~Sun, L.~Mitchell, K.~Kulkarni, A.~Klöckner, D.~A. Ham, and P.~H.~J.
  Kelly}, {\em A study of vectorization for matrix-free finite element
  methods}, Int.~J.~of~High~Perf.~Comp.~App.,  (2019),
  \url{https://doi.org/10.1177/1094342020945005}.

\bibitem{Thoennes:2023:ModelBasePerfAnalysisHyteg}
{\sc D.~Th\"{o}nnes and U.~R\"{u}de}, {\em Model-based performance analysis of
  the {HyTeG} finite element framework}, in Proceedings of the Platform for
  Advanced Scientific Computing Conference, PASC '23, New York, NY, USA, 2023,
  Association for Computing Machinery,
  \url{https://doi.org/10.1145/3592979.3593422}.

\bibitem{Wellein:2010:Likwid}
{\sc J.~Treibig, G.~Hager, and G.~Wellein}, {\em {LIKWID}: A lightweight
  performance-oriented tool suite for x86 multicore environments}, in
  Proceedings of PSTI2010, the First International Workshop on Parallel
  Software Tools and Tool Infrastructures, San Diego CA, 2010,
  \url{https://doi.org/10.1109/ICPPW.2010.38}.

\bibitem{Williams:2009:Roofline}
{\sc S.~Williams, A.~Waterman, and D.~Patterson}, {\em Roofline: An insightful
  visual performance model for multicore architectures}, Commun. ACM, 52
  (2009), pp.~65--76, \url{https://doi.org/10.1145/1498765.1498785}.

\bibitem{Xiao:2010:Quad}
{\sc H.~Xiao and Z.~Gimbutas}, {\em A numerical algorithm for the construction
  of efficient quadrature rules in two and higher dimensions}, Computers and
  Mathematics with Applications, 59 (2010), pp.~663--676,
  \url{https://doi.org/10.1016/j.camwa.2009.10.027}.

\bibitem{Zhong:2009:OptimalErrorEstimates}
{\sc L.~Zhong, S.~Shu, G.~Wittum, and J.~Xu}, {\em Optimal error estimates for
  {N}\'{e}d\'{e}lec edge elements for time-harmonic {M}axwell's equations},
  J.~Comp.~Math., 27 (2009), pp.~563--572,
  \url{https://doi.org/10.4208/jcm.2009.27.5.011}.

\bibitem{Olgaard:2010:OptimQuadRepr}
{\sc K.~B. Ølgaard and G.~N. Wells}, {\em Optimizations for quadrature
  representations of finite element tensors through automated code generation},
  ACM Transactions on Mathematical Software, 37 (2010),
  \url{https://doi.org/10.1145/1644001.1644009}.

\end{thebibliography}

\end{document}